\documentclass[useAMS, usenatbib, usegraphicx]{mn2e}
\usepackage{dcolumn}
\usepackage{bm}
\usepackage{amssymb}
\usepackage{color}

\newcommand{\m}{\phantom{-}}
\newcommand{\z}{\phantom{0}}
\newcommand{\zz}{\phantom{00}}
\newcommand{\zzzp}{\phantom{(000)}}

\newcommand{\lowentry}[1]{\smash{\lower 1.5 ex \hbox{#1}}}
\newcommand{\highentry}[1]{\smash{\raise 1.5 ex \hbox{#1}}}

\begin{document}

\title[Dynamic migration of rotating neutron stars]
      {Dynamic migration of rotating neutron stars due to a phase
        transition instability}

\author[H.~Dimmelmeier, M.~Bejger, P.~Haensel and J.~L.~Zdunik]
       {Harald~Dimmelmeier$^{1}$\thanks{E-mail:harrydee@mpa-garching.mpg.de},
        Michal~Bejger$^{2}$\thanks{E-mail:bejger@camk.edu.pl},
        Pawel~Haensel$^{2}$\thanks{E-mail:haensel@camk.edu.pl} and
        J.~Leszek~Zdunik$^{2}$\thanks{E-mail:jlz@camk.edu.pl} \vspace*{0.2 em} \\
        $^1$Department of Physics, Aristotle University of
        Thessaloniki, GR-54124 Thessaloniki, Greece \\
        $^2$Copernicus Astronomical Center, Polish Academy of
        Sciences, Bartycka 18, 00-716 Warszawa, Poland}

\date{Accepted $<$date$>$. Received $<$date$>$; in original form $<$date$>$}

\pagerange{\pageref{firstpage}--\pageref{lastpage}} \pubyear{2008}

\label{firstpage}

\maketitle

%%%%%%%%%%%%%%%%%%%%%%%%%%%%%%%%%%%%%%%%%%%%%%%%%%%%%%%%%%%%%%%%%%%%%%%%%%%%%%%%%%%
%%%%%%%%%%%%%%%%%%%%%%%%%%%%%%%%%%%%%%%%%%%%%%%%%%%%%%%%%%%%%%%%%%%%%%%%%%%%%%%%%%%
%%%%%%%%%%%%%%%%%%%%%%%%%%%%%%%%%%%%%%%%%%%%%%%%%%%%%%%%%%%%%%%%%%%%%%%%%%%%%%%%%%%

\begin{abstract}
  Using numerical simulations based on solving the general
  relativistic hydrodynamic equations with the CoCoNuT code,
  we study the dynamics of a phase transition in the dense core of
  isolated rotating neutron stars, triggered by the back bending
  instability reached via angular momentum loss. In particular, we
  investigate the dynamics of a migration from an unstable
  configuration into a stable one, which leads to a mini-collapse of
  the neutron star and excites sizeable pulsations in its bulk until
  it acquires a new stable equilibrium state. We consider two
  equations of state which exhibit softening at high densities, a
  simple analytic one with a mixed hadron-quark phase (where the
  hadron pressure is approximated by a polytrope) in an intermediate
  pressure interval and pure quark matter at very high densities, and
  a microphysical one that has a first-order phase transition at
  constant pressure with a jump in density, originating from kaon
  condensation. Although the marginally stable initial models are
  rigidly rotating, we observe that during the collapse (albeit
  little) differential rotation is created. We analyze the emission of
  gravitational radiation in such an event, which in some models is
  amplified by mode resonance effects, and assess its prospective
  detectability by current and future interferometric detectors. We
  expect that the  most favorable conditions for dynamic migration
  exist in very young magnetars. The rate of such events in our Galaxy
  is of the order of one per century and rises to about one per year
  if the Virgo cluster of galaxies is considered. We find that the
  damping of the post-migration pulsations and, accordingly, of the
  gravitational wave signal amplitude strongly depends on the
  character of the equation of state softening (either via a density
  jump or continuous through a mixed state). The damping of pulsations
  in the models with the microphysical equation of state is caused by
  dissipation associated with matter flowing through the density jump
  at the edge of the dense core. If at work, this mechanism dominates
  over all other types of dissipation, like bulk viscosity in the
  exotic-phase core, gravitational radiation damping, or numerical
  viscosity.
\end{abstract}

\begin{keywords}
  hydrodynamics -- relativity -- methods: numerical -- stars: neutron --
  stars: pulsations -- stars: phase-transition -- stars: rotation
\end{keywords}

%%%%%%%%%%%%%%%%%%%%%%%%%%%%%%%%%%%%%%%%%%%%%%%%%%%%%%%%%%%%%%%%%%%%%%%%%%%%%%%%%%%
%%%%%%%%%%%%%%%%%%%%%%%%%%%%%%%%%%%%%%%%%%%%%%%%%%%%%%%%%%%%%%%%%%%%%%%%%%%%%%%%%%%
%%%%%%%%%%%%%%%%%%%%%%%%%%%%%%%%%%%%%%%%%%%%%%%%%%%%%%%%%%%%%%%%%%%%%%%%%%%%%%%%%%%

\section{Introduction}
\label{section:introduction}

One of the mysteries of neutron stars is the structure of their
cores where the density exceeds normal nuclear density
$ \rho_\mathrm{nuc} \approx 3 \times 10^{14} \mathrm{\ g\ cm}^{-3} $.
For $ \rho \lesssim \rho_\mathrm{nuc} $ the constituents of dense
matter are well known (neutrons, protons, and electrons). Here the
respective physical theories can be confronted with nuclear physics
data and are therefore under control \citep[see e.g.][]{haensel_07_a}.
However, for $ \rho $ significantly larger than $ \rho_\mathrm{nuc} $,
the predictions of different  plausible theories diverge with
increasing density. The structure of matter at
$ 5\mbox{\,--\,}10 \, \rho_\mathrm{nuc} $, i.e.\ at densities
characteristic for the center of massive neutron stars, is unknown.
Still, in order to construct neutron star models, the equation of
state (EoS) of matter in this regime is needed as an input. Some
theories of dense matter predict a strong softening of the EoS at
$ \rho \gtrsim 2 \, \rho_\mathrm{nuc} $, caused by the appearance
of various exotic phases of matter, such as pion condensate, kaon
condensate, or deconfined quark matter \citep[for a review
see][]{haensel_07_a, glendenning_00_a, weber_99_a}. The prediction of
observable signatures of the presence of exotic phases in the dense
core of neutron stars is thus crucial for testing physical theories of
hadronic matter at extreme densities.

A softening of the EoS is expected to influence the rotational
properties of neutron stars. Today, we know some 1800 radio pulsars
(i.e.\ rotating magnetized neutron stars), which are spinning down due
to angular momentum loss via radiation. They are observed using radio
telescopes, where the evolution of the rotation period with time is
measured, and some of them are also being monitored by the newly
constructed gravitational wave detectors (albeit without a reliable
detection so far in that case). If the EoS of matter in a pulsar core
actually softens at some density threshold, this could have a dramatic
impact on the structure of the neutron star that completely alters the
pulsar timing properties: One of the possibilities is that the pulsar
could enter a back bending episode, which is reflected by the reversal
of the spin-down into a spin-up \citep{glendenning_97_a, zdunik_06_a}.
If the softening of the EoS is sufficiently strong, the back bending
could lead to an instability at which the rotating neutron star
dynamically migrates to a new stable equilibrium state
\citep{zdunik_06_a}.

Depending on the mass of the neutron star, this migration process can
be accompanied by a considerable increase in central compactness and
spin-up. Immediately after migration, the spinning neutron star's bulk
is expected to pulsate with sizeable amplitude, which would result in
a burst of gravitational radiation. This prospect gives additional
motivation for studying the dynamical migration phenomenon. A scenario
with similar dynamics (i.e.\ a dynamic migration of rotating neutron
stars towards a new stable equilibrium state at higher central density
along with a spin-up, the excitation of strong pulsations, and the
emission of gravitational waves) was recently investigated by
\citet{lin_06_a} and \citet{abdikamalov_09_a}. However, while in that
work the original, unperturbed initial configuration is in stable
equilibrium and the transient is initated by an instantaneous change
in the EoS (resulting in a somewhat artificial depletion of pressure
at high densities), the migration process due to back bending need not
be initiated by an ad-hoc modification of the EoS, as this instability
starts at an unstable equilibrium state.

In this work we investigate the dynamics of such a migration of pulsar
models, triggered by the instability described above. This is
accomplished by performing fully nonlinear time-dependent numerical
simulations with a two-dimensional version of the general relativistic
hydrodynamics code CoCoNuT. From a multitude of existing EoSs
we select two types which feature softening at high densities: The
first is the simple analytic MUn EoS of \citet{zdunik_06_a} which
obeys a polytropic relation at low densities, describes a mixed
hadron-quark phase at intermediate densities (where the hadron
pressure is again approximated by a polytrope, however with a
different adiabatic exponent), and has pure quark matter at very high
densities. The second one is a tabulated microphysical EoS that
exhibits a first-order phase transition at constant pressure with a
jump in density, originating from kaon condensation. For these two
EoSs, as initial models we select marginally stable rotating
configurations with different baryon masses, and follow the migration
and the subsequent ring-down phase for many dynamic time scales. We
analyze the post-migration pulsations in terms of amplitude, spectrum
and possible damping mechanisms. Furthermore, we calculate the
associated gravitational wave emission and discuss the prospects for a
possible detection by current and future interferometric detectors
like LIGO, VIRGO or Advanced LIGO.

This article is organized as follows: In
Section~\ref{section:formulation_and_numerics} we present the
mathematical framework and numerical setup for the simulations, while
in Section~\ref{section:models} we give details about the two EoSs
employed and introduce the marginally stable models that undergo
migration. Section~\ref{section:results} contains a survey of the
results, with a general description of the migration dynamics, an
analysis of the pulsations excited by this process, information about
the final equilibrium states, followed by a discussion of our models'
gravitational wave emission and the detectability prospects. In
Section~\ref{section:conclusions} we summarize our findings and close
with some conclusions. We also add some information about
non-equilibrium effects in the EoSs in
Appendix~\ref{appendix:nonequilibrium_effects} and explain the damping
mechanism for pulsations in models with the microphysical EoS in
Appendix~\ref{appendix:pulsation_damping}.

Unless otherwise noted, we choose dimensionless units for all physical
quantities by setting the speed of light and the gravitational
constant to one, $ c = G = 1 $. Latin indices run from 1 to 3, Greek
indices from 1 to 4.

%%%%%%%%%%%%%%%%%%%%%%%%%%%%%%%%%%%%%%%%%%%%%%%%%%%%%%%%%%%%%%%%%%%%%%%%%%%%%%%%%%%
%%%%%%%%%%%%%%%%%%%%%%%%%%%%%%%%%%%%%%%%%%%%%%%%%%%%%%%%%%%%%%%%%%%%%%%%%%%%%%%%%%%
%%%%%%%%%%%%%%%%%%%%%%%%%%%%%%%%%%%%%%%%%%%%%%%%%%%%%%%%%%%%%%%%%%%%%%%%%%%%%%%%%%%

\section{Formulation and numerical methods}
\label{section:formulation_and_numerics}

We construct the initial equilibrium models of rotating neutron stars
using a variant of the self-consistent field method described in
\citet{komatsu_89_a} (KEH hereafter), as implemented in the
code \textsc{RNS} \citep{stergioulas_95_a}. This code solves the
general relativistic hydrostationary equations for rotating matter
distributions whose pressure obeys a barotropic EoS. The resulting
equilibrium models are taken as initial data for the evolution code.

The time dependent numerical simulations are performed with the code
CoCoNuT developed by \citet{dimmelmeier_02_a,
dimmelmeier_02_b} with a metric solver based on spectral methods as
described in \citet{dimmelmeier_05_a}. The code solves the general
relativistic field equations for a curved spacetime in the $ 3 + 1 $
split under the assumption of the conformal flatness condition (CFC)
for the three-metric. The hydrodynamics equations are formulated in
conservation form, and are evolved by high-resolution shock-capturing
schemes.

In the following, we present the mathematical formulation of the
metric and hydrodynamics equations, and then summarize the numerical
methods used for solving them. Note that this part is practically
identical to Section~2 in \citet{abdikamalov_09_a}.

%%%%%%%%%%%%%%%%%%%%%%%%%%%%%%%%%%%%%%%%%%%%%%%%%%%%%%%%%%%%%%%%%%%%%%%%%%%%%%%%%%%
%%%%%%%%%%%%%%%%%%%%%%%%%%%%%%%%%%%%%%%%%%%%%%%%%%%%%%%%%%%%%%%%%%%%%%%%%%%%%%%%%%%

\subsection{Metric equations}
\label{subsection:metric_equations}

We adopt the ADM 3\,+\,1 formalism by \citet{arnowitt_62_a} to
foliate a spacetime endowed with a metric $ g_{\mu\nu} $ into a set of
non-intersecting spacelike hypersurfaces. The line element then reads
\begin{equation}
  ds^2 = g_{\mu\nu} \, dx^\mu \, dx^\nu = - \alpha^2 dt^2 +
  \gamma_{ij} (dx^i + \beta^i dt) (dx^j + \beta^j dt), ~
  \label{eq:line_element}
\end{equation}
where $ \alpha $ is the lapse function, $ \beta^i $ is the spacelike
shift three-vector, and $ \gamma_{ij} $ is the spatial three-metric.

In the 3\,+\,1 formalism, the Einstein equations are split into
evolution equations for the three-metric $ \gamma_{ij} $ and the
extrinsic curvature $ K_{ij} $, and constraint equations (the
Hamiltonian and momentum constraints) which must be fulfilled at every
spacelike hypersurface.

The fluid is specified by the rest-mass density $ \rho $, the
four-velocity $ u^\mu $, and the pressure $ P $, with the specific
enthalpy defined as $ h = 1 + \epsilon + P / \rho $, where
$ \epsilon $ is the specific internal energy. The three-velocity of
the fluid as measured by an Eulerian observer is given by
$ v^i = u^i / (\alpha u^0) + \beta^i / \alpha $, and the Lorentz
factor $ W = \alpha u^0 $ satisfies the relation
$ W = 1 / \sqrt{1 - v_i v^i} $.

Based on the ideas of \citet{isenberg_08_a} and \citet{wilson_96_a},
and as it was done in the work of \citet{dimmelmeier_02_a,
  dimmelmeier_02_b}, we approximate the general metric $ g_{\mu\nu} $
by replacing the spatial three-metric $ \gamma_{ij} $ with the
conformally flat three-metric
\begin{equation}
  \gamma_{ij} = \phi^4 \hat{\gamma}_{ij},
  \label{eq:cfc_metric}
\end{equation}
where $ \hat{\gamma}_{ij} $ is the flat metric and $ \phi $ is a
conformal factor. In this CFC approximation, the ADM equations for the
spacetime metric reduce to a set of five coupled elliptic non-linear
equations for the metric components,
\begin{equation}
  \setlength{\arraycolsep}{0.14 em}
  \begin{array}{rcl}
    \hat{\Delta} \phi & = & \displaystyle - 2 \pi \phi^5
    \left( \rho h W^2 - P \right) - \phi^5 \frac{K_{ij} K^{ij}}{8},
    \\ [1.2 em]
    \hat{\Delta} (\alpha \phi) & = & \displaystyle 2 \pi \alpha \phi^5
    \left( \rho h (3 W^2 - 2) + 5 P \right) +
    \alpha \phi^5 \frac{7 K_{ij} K^{ij}}{8},
    \\ [1.2 em]
    \hat{\Delta} \beta^i & = & \displaystyle 16 \pi \alpha \phi^4
    \rho h W^2 v^i + 2 \phi^{10} K^{ij} \hat{\nabla}_{\!j}
    \! \left( \alpha \phi^{-6} \right) -
    \frac{1}{3} \hat{\nabla}^i \hat{\nabla}_{\!k} \beta^k\!,
    \!\!\!\!\!\!\!\!\!\!\!\!\!\!\!\!\!\!\!\!\!
  \end{array}
  \label{eq:cfc_metric_equations}
\end{equation}
where the maximal slicing condition, $ K^i_i = 0 $, is imposed. Here
$ \hat{\nabla}_{\!i} $ and $ \hat{\Delta} $ are the flat space Nabla
and Laplace operators, respectively. For the extrinsic curvature we
have the expression
\begin{equation}
  K_{ij} = \frac {1}{2 \alpha}
  \left( \! \nabla_{\!i} \beta_j + \nabla_{\!j} \beta_i -
  \frac{2}{3} \gamma_{ij} \nabla_{\!k} \beta^k \! \right),
  \label{eq:definition_of_extrinsic_curvature}
\end{equation}
which closes the system~(\ref{eq:cfc_metric_equations}).

We rewrite the above metric equations in a mathematically equivalent
form by introducing an auxiliary vector field $ W^i $ and obtain
\citep{saijo_04_a}
\begin{equation}
  \setlength{\arraycolsep}{0.14 em}
  \begin{array}{rcl}
    \hat{\Delta} \phi & = & \displaystyle - 2 \pi \phi^5
    \left( \rho h W^2 - P \right) -
    \phi^{-7} \frac{\hat{K}_{ij} \hat{K}^{ij}}{8},
    \\ [1.2 em]
    \hat{\Delta} (\alpha \phi) & = & \displaystyle 2 \pi \alpha \phi^5
    \left( \rho h (3 W^2 - 2) + 5 P \right) +
    \alpha \phi^{-7} \frac{7 \hat{K}_{ij} \hat{K}^{ij}}{8},
    \\ [1.2 em]
    \hat{\Delta} \beta^i & = & \displaystyle 2 \hat{\nabla}_{\!j} \!
    \left( 2 \alpha \phi^{-6} \hat{K}^{ij} \right) -
    \frac{1}{3} \hat{\nabla}^i \hat{\nabla}_{\!k} \beta^k\!,
    \\ [1.2 em]
    \hat{\Delta} W^i & = & \displaystyle 8 \pi \phi^{10} \rho h W^2 v^i -
    \frac{1}{3} \hat{\nabla}^i \hat{\nabla}_{\!k} W^k\!,
  \end{array}
  \label{eq:cfc_metric_equations_new}
\end{equation}
where the flat space extrinsic curvature is given by
\begin{equation}
  \hat{K}_{ij} = \hat{\nabla}_{\!i} W_j + \hat{\nabla}_{\!j} W_i -
  \frac{2}{3} \hat{\gamma}_{ij} \hat{\nabla}_{\!k} W^k
  \label{eq:definition_of_flat_extrinsic_curvature}
\end{equation}
and relates to the regular extrinsic curvature as
$ \hat{K}_{ij} = \phi^2 K_{ij} $ and
$ \hat{K}^{ij} = \phi^{10} K^{ij} $. The virtue of this reformulation
of the metric equations is discussed in detail by \citet{cordero_09_a}.

Note that the metric equations do not contain explicit time
derivatives, and thus the metric is calculated by a fully constrained
approach, at the cost of neglecting some evolutionary degrees of
freedom in the spacetime metric (e.g.\ dynamic gravitational wave
degrees of freedom).

The accuracy of the CFC approximation has been tested in various
works, both in the context of stellar core collapse and for
equilibrium models of neutron stars \citep[for a detailed comparison
between the CFC approximation and full general relativity, see][and
references therein]{dimmelmeier_06_b, ott_07_a}. The spacetime of
rapidly (uniformly or differentially) rotating neutron star models is
still very well approximated by the CFC metric~(\ref{eq:cfc_metric}).
The accuracy of the approximation is expected to degrade only in
extreme cases, such as a rapidly rotating black hole or compact binary
systems.

%%%%%%%%%%%%%%%%%%%%%%%%%%%%%%%%%%%%%%%%%%%%%%%%%%%%%%%%%%%%%%%%%%%%%%%%%%%%%%%%%%%
%%%%%%%%%%%%%%%%%%%%%%%%%%%%%%%%%%%%%%%%%%%%%%%%%%%%%%%%%%%%%%%%%%%%%%%%%%%%%%%%%%%

\subsection{General relativistic hydrodynamics}
\label{subsection:gr_hydrodynamics}

The hydrodynamic evolution of a relativistic perfect fluid is
determined by a system of local conservation equations, which read
\begin{equation}
  \nabla_{\!\mu} J^{\mu} = 0, \qquad \nabla_{\!\mu} T^{\mu \nu} = 0,
  \label{eq:gr_equations_of_motion}
\end{equation}
where $ J^{\mu} = \rho u^{\mu} $ is the rest-mass current, and
$\nabla_{\!\mu}$ denotes the covariant derivative with respect to the
four-metric $ g_{\mu \nu} $. Following \citet{banyuls_97_a} we
introduce a set of conserved variables in terms of the primitive
(physical) variables $ (\rho, v_i, \epsilon) $:
\begin{equation}
  D = \rho W,
  \qquad
  S_i = \rho h W^2 v_i,
  \qquad
  \tau = \rho h W^2 - P - D.
  \label{eq:conserved_quantities}
\end{equation}
Using the above variables, the local conservation
laws~(\ref{eq:gr_equations_of_motion}) can be written as a
first-order, flux-conservative hyperbolic system of equations,
\begin{equation}
  \frac{\partial \sqrt{\gamma} \, \bm{U}}{\partial t} +
  \frac{\partial \sqrt{- g} \bm{F}^i}{\partial x^i} =
  \sqrt{- g} \bm{S},
  \label{eq:hydro_conservation_equation}
\end{equation}
with the state vector, flux vector, and source vector
\begin{equation}
  \setlength{\arraycolsep}{0.14 em}
  \begin{array}{rcl}
  \bm{U} & = & [D, S_j, \tau], \\ [1.0 em]
  \bm{F}^i & = & \displaystyle
  \left[ D \hat{v}^i, S_j \hat{v}^i + \delta^i_j P,
  \tau \hat{v}^i + P v^i \right], \\ [1.0 em]
  \bm{S} & = & \displaystyle
  \Bigl[ 0, \frac{1}{2} T^{\mu \nu}
  \frac{\partial g_{\mu \nu}}{\partial x^j},
  T^{00} \left( K_{ij} \beta^i \beta^j - \beta^j
  \frac{\partial \alpha}{\partial x^j} \right) + \\ [1.0 em]
  & & \displaystyle \qquad \qquad \qquad \;\:\:
  T^{0j} \left( 2 K_{ij} \beta^i -
  \frac{\partial \alpha}{\partial x^j} \right) +
  T^{ij} K_{ij} \Bigr],
  \end{array}
  \label{eq:hydro_conservation_equation_constituents}
\end{equation}
respectively. Here $ \hat{v}^i = v^i - \beta^i / \alpha $, and
$ \sqrt{-g} = \alpha \sqrt{\gamma} $, with $ g = \det (g_{\mu \nu}) $
and $ \gamma = \det (\gamma_{ij}) $.

The system of hydrodynamics
equations~(\ref{eq:hydro_conservation_equation}) is closed by an
EoS, which relates the pressure to some thermodynamically independent
quantities, in our case $ P = P (\rho, \epsilon) $.

%%%%%%%%%%%%%%%%%%%%%%%%%%%%%%%%%%%%%%%%%%%%%%%%%%%%%%%%%%%%%%%%%%%%%%%%%%%%%%%%%%%
%%%%%%%%%%%%%%%%%%%%%%%%%%%%%%%%%%%%%%%%%%%%%%%%%%%%%%%%%%%%%%%%%%%%%%%%%%%%%%%%%%%

\subsection{Numerical methods for solving the metric and hydrodynamics
  equations}
\label{subsection:numerical_methods}

The hydrodynamic solver performs the numerical time integration of the
system of conservation
equations~(\ref{eq:hydro_conservation_equation}) using a
high-resolution shock-capturing (HRSC) scheme on a finite-difference
grid. In (upwind) HRSC methods a Riemann problem has to be solved at
each cell interface, which requires the reconstruction of the
primitive variables $ (\rho, v^i, \epsilon) $ at these interfaces. We
use the piecewise parabolic method (PPM) method for the
reconstruction, which yields third-order accuracy in space for smooth
flows and away from extrema. The numerical fluxes are computed by
means of Marquina's approximate flux formula \citep{donat_98_a}. The
time update of the conserved vector $ \bm{U} $ is done using the
method of lines in combination with a Runge--Kutta scheme with
second-order accuracy in time. Once the state vector is updated in
time, the primitive variables are recover through an iterative
Newton--Raphson method. To numerically solve the elliptic CFC metric
equations~(\ref{eq:cfc_metric_equations}) we make use of an iterative
non-linear solver based on spectral methods. The combination of HRSC
methods for the hydrodynamics and spectral methods for the metric
equations in a multidimensional numerical code has been in detail in
\citet{dimmelmeier_05_a}.

The CoCoNuT code utilizes Eulerian spherical polar
coordinates $ \{r, \theta\} $, as for the models discussed in this
work we assume axisymmetry with respect to the rotation axis and
additionally equatorial symmetry. The finite-difference grid consists
of 160 radial and 40 angular grid points, which are equidistantly
spaced. A small part of the grid, which initially corresponds to 20
radial grid points, covers an artificial low-density atmosphere
extending beyond the stellar surface, whose rest-mass density is
$ 10^{-17} $ of the initial central rest-mass density of the star.

Since the calculation of the spacetime metric is computationally
expensive, the metric is updated only once every 25 hydrodynamic
time steps during the evolution and extrapolated when needed. The
suitability of this procedure is tested and discussed in detail
in \citet{dimmelmeier_02_a}. We also note that tests with different
grid resolutions were performed to ascertain that the regular grid
resolution specified above is appropriate for our
simulations.

%%%%%%%%%%%%%%%%%%%%%%%%%%%%%%%%%%%%%%%%%%%%%%%%%%%%%%%%%%%%%%%%%%%%%%%%%%%%%%%%%%%
%%%%%%%%%%%%%%%%%%%%%%%%%%%%%%%%%%%%%%%%%%%%%%%%%%%%%%%%%%%%%%%%%%%%%%%%%%%%%%%%%%%

\subsection{Gravitational waves}
\label{subsection:gravitational_waves}

The gravitational waves emitted by the collapsing neutron star are
computed using the Newtonian quadrupole formula in its first
time-integrated form \citep[the first-moment of momentum density
formulation as described in detail in][]{dimmelmeier_02_b} in the
variant of \citet{shibata_04_a}. It yields the quadrupole wave
amplitude $ A^\mathrm{E2}_{20} $ as the lowest order term in a
multipole expansion of the radiation field into pure-spin tensor
harmonics \citep{thorne_80_a}. The wave amplitude is related to the
dimensionless gravitational wave strain $ h $ in the equatorial plane
\citep{dimmelmeier_02_b}
\begin{equation}
  h = \frac{1}{8} \sqrt{\frac{15}{\pi}} \frac{A^\mathrm{E2}_{20}}{r} =
  8.8524 \times 10^{-21} \frac{A^\mathrm{E2}_{20}}{10^3 \mathrm{\ cm}}
  \frac{10 \mathrm{\ kpc}}{r},
\end{equation}
with $ r $ being the distance to the emitting source.

We point out that although the quadrupole formula is not gauge
invariant and is only strictly valid in the Newtonian slow-motion
limit, for gravitational waves emitted by pulsations of rotating
neutron stars it yields results that agree very well in phase and to
about $ 10\%\mbox{\,--\,}20\% $ in amplitude with more sophisticated
methods \citep{shibata_03_a, nagar_07_a}.

%%%%%%%%%%%%%%%%%%%%%%%%%%%%%%%%%%%%%%%%%%%%%%%%%%%%%%%%%%%%%%%%%%%%%%%%%%%%%%%%%%%
%%%%%%%%%%%%%%%%%%%%%%%%%%%%%%%%%%%%%%%%%%%%%%%%%%%%%%%%%%%%%%%%%%%%%%%%%%%%%%%%%%%
%%%%%%%%%%%%%%%%%%%%%%%%%%%%%%%%%%%%%%%%%%%%%%%%%%%%%%%%%%%%%%%%%%%%%%%%%%%%%%%%%%%

\section{Model setup}
\label{section:models}

%%%%%%%%%%%%%%%%%%%%%%%%%%%%%%%%%%%%%%%%%%%%%%%%%%%%%%%%%%%%%%%%%%%%%%%%%%%%%%%%%%%
%%%%%%%%%%%%%%%%%%%%%%%%%%%%%%%%%%%%%%%%%%%%%%%%%%%%%%%%%%%%%%%%%%%%%%%%%%%%%%%%%%%

\subsection{Equations of state}
\label{subsection:equations_of_state}

One of our two sets of models is based on a simple analytic
three-phase EoS which is identical to the mixed unstable (MUn) EoS
discussed in \citet{zdunik_06_a}\footnote{Note that here we give the
  EoS in terms of the rest-mass density $ \rho $ (the quantity used by
  our evolution code) rather than the baryon number density
  $ n = \rho / m_\mathrm{u} $, where
  $ m_\mathrm{u} = 1.66 \times 10^{-24} \mathrm{g} $ is the baryon
  mass.}:

\begin{itemize}
\item In the normal hadronic phase, i.e.\ for
  $ \rho < \rho_1 = 3.32 \times 10^{14} \mathrm{\ g\ cm}^{-3} $
  (corresponding to $ n_1 = 0.2 \times 10^{39} \mathrm{cm}^{-3} $),
  the EoS is given by a polytrope,
  \begin{equation}
    P = K \rho^\gamma,
    \label{eq:polytropic_eos}
  \end{equation}
  with $ \gamma = \gamma_\mathrm{h} = 2.5 $ and a polytropic constant
  $ K = K_\mathrm{h} = 1.0506 \times 10^{-2} $ \citep[in cgs units,
  which equals $ 0.025 $ in units of
  $ \hat{\rho} c^2 / \hat{n}^{\gamma_\mathrm{h}} $ with
  $ \hat{\rho} = 1.66 \times 10^{14} \mathrm{\ g\ cm}^{-3} $ and
  $ \hat{n} = 0.1 \times 10^{39} \mathrm{cm}^{-3} $,
  see][]{zdunik_06_a}. The internal energy is calculated according to
  the ideal gas EoS as
  \begin{equation}
    \epsilon = \frac{P}{\rho (\gamma - 1)}.
    \label{eq:ideal_gas_eos}
  \end{equation}
\item In the mixed phase, i.e.\ for
  $ \rho_1 \le \rho < \rho_2 = 10.79 \times 10^{14} \mathrm{\ g\ cm}^{-3} $
  (corresponding to $ n_2 = 0.65 \times 10^{39} \mathrm{cm}^{-3} $),
  the EoS is again a polytrope, now with
  $ \gamma = \gamma_\mathrm{m} = 1.3 $ and a polytropic constant
  $ K = K_\mathrm{m} = 2.7976 \times 10^{15} $ \citep{zdunik_00_a},
  which ensures continuity of $ P $ at $ \rho_1 $. The expression for
  the internal energy is more complicated than in the regular phase,
  \begin{equation}
    \epsilon = \frac{P}{\rho (\gamma - 1)} +
    \frac{m_\mathrm{m}}{m_\mathrm{u}} - 1,
    \label{eq:internal_energy_mixed_phase}
  \end{equation}
  where the \emph{mean} atomic mass
  \begin{equation}
    m_\mathrm{m} = m_\mathrm{u} \left( 1 - \frac{P_1}{\rho_1}
    \frac{\gamma_\mathrm{h} - \gamma_\mathrm{m}}
    {(\gamma_\mathrm{h} - 1) (\gamma_\mathrm{m} - 1)} \right).
    \label{eq:fiducial_mixed_phase_atomic_mass}
  \end{equation}
  in the mixed phase is determined by demanding that also $ \epsilon $
  is continuous at $ \rho_1 $. Here $ P_1 $ is the pressure at
  $ \rho_1 $.
\item At high densities $ \rho \ge \rho_2 $, we assume a pure quark
  matter phase, whose pressure is given by the MIT bag model
  \citep{zdunik_00_a},
  \begin{equation}
    P = \frac{1}{3} (\mathcal{E} - \mathcal{E}_2) + P_2,
    \qquad
    \rho = \rho_2 \left( \frac{4 P + \mathcal{E}_2 - 3 P_2}
    {\mathcal{E}_2 + P_2} \right)^{3/4}\!\!\!\!\!\!\!\!,
    \label{eq:quark_matter}
  \end{equation}
  where $ \mathcal{E} = \rho (1 + \epsilon) $ is the total energy
  density of the quark matter\footnote{By assuming that
    $ \epsilon = \mathcal{E} / \rho - 1 $ we can continue to use
    $ \rho $ and $ \epsilon $ in our system of hydrodynamic
    equations~(\ref{eq:hydro_conservation_equation}); for a more
    detailed discussion of this issue, see \citet{abdikamalov_09_a}.}
  and $ P_2 $ and $ \mathcal{E}_2 $ are the pressure and energy
  density at $ \rho_2 $, respectively, which again enforces continuity
  of the EoS at the transition point.
\end{itemize}

The other set of models uses an microphysical EoS, which exhibits a
phase transition to a kaon condensate accompanied by a density
discontinuity. The composition of this EoS is as follows:

\begin{itemize}
\item For the description of the inner crust (nuclei immersed in
  neutron and electron gas) and the outer crust (nuclei immersed in
  electron gas) at densities $ 5 \times 10^{10} \mathrm{\ g\ cm}^{-3}
  \le \rho \le 1.5 \times 10^{14} \mathrm{\ g\ cm}^{-3} $ we use the
  model of \cite{douchin_01_a}. In this model, neutron drip occurs at
  a density of $ 3 \times 10^{11} \mathrm{\ g\ cm}^{-3} $. The outer
  crust at densities $ 10^{8} \mathrm{\ g\ cm}^{-3} \le \rho \le
  5 \times 10^{10} \mathrm{\ g\ cm}^{-3} $ is modeled by the EoS of
  \citet{haensel_94_a}. Finally, the low-density envelope with
  $ \rho < 10^{8} \mathrm{\ g\ cm}^{-3} $ is described by the EoS of
  \citet{baym_71_a}.
\item The constituents of the moderately dense part of the core below
  the lower end of the density jump at
  $ \rho_1 = 6.31 \times 10^{14} \mathrm{\ g\ cm}^{-3} $ are neutrons,
  protons, electrons, and muons. The nucleon component is described by
  a relativistic mean-field model with scalar self-coupling, as
  constructed by \citet{zimanyi_90_a}. The values of the
  meson--nucleon coupling constants are
  $ g_\sigma / m_\sigma = 3.122 \mathrm{\ fm} $,
  $ g_\omega / m_\omega = 2.1954 \mathrm{\ fm} $, and
  $ g_\rho / m_\rho = 2.1888 \mathrm{\ fm} $. The dimensionless
  coefficients in the cubic and quartic terms in the scalar
  self-coupling are $ b = - 6.418 \times 10^{-3} $ and
  $ c = 2.968 \times 10^{-3} $, respectively.
\item Although ideally the pressure $ P $ should remain constant
  across the density jump from $ \rho_1 $ to
  $ \rho_2 = 10.24 \times 10^{14} \mathrm{\ g\ cm}^{-3} $, in order to
  avoid numerical problems with the hydrodynamics scheme of the
  CoCoNuT code we introduce a small linear pressure increase
  of 10\% in this density interval.
\item For densities higher than the upper end of the density jump at
  $ \rho_2 $ we consider kaon-condensed matter. The coupling of kaons
  to nucleons is done according to the model of
  \citet{glendenning_99_a} with
  $ U_K^\mathrm{lin} = - 110 \mathrm{\ MeV} $.
\end{itemize}

\begin{figure}
  \centerline{\includegraphics[width = 85 mm]{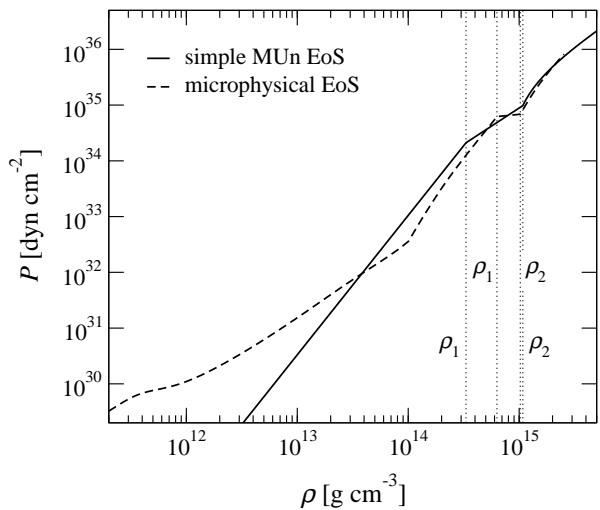}}
  \caption{Dependence of the pressure $ P $ on the density $ \rho $
    for the simple MUn EoS (solid line) and the microphysical EoS
    (dashed line). The transition densities $ \rho_1 $ and $ \rho_2 $
    for both the MUn EoS and the microphysical EoS are indicated by
    the dotted lines. Note the domain of zero pressure slope in the
    microphysical EoS.}
  \label{fig:equation_of_state}
\end{figure}

The properties of these two EoSs are exemplified in
Fig.~\ref{fig:equation_of_state}, where we plot the dependence of the
pressure $ P $ on the rest-mass density $ \rho $. In the case of the
microphysical EoS the approximately constant pressure between
$ \rho_1 $ and $ \rho_2 $ is apparent. In the dense phase, both EoSs
have a very similar shape.

Our approach assumes that the fluid is in thermodynamic equilibrium
everywhere, even during the dynamic migration process. The
justification of this assumption is given in
Appendix~\ref{appendix:nonequilibrium_effects}.

As both the simple MUn EoS and the microphyical EoS are barotropic,
there is no need to evolve the quantity $ \tau $ in the hyperbolic set
of hydrodynamic evolution
equations~(\ref{eq:hydro_conservation_equation},
\ref{eq:hydro_conservation_equation_constituents}). During the
recovery of the primitive variables, the associated quantity
$ \epsilon $ is simply evaluated from the EoS using the rest-mass
density $ \rho $, and then $ \tau $ is computed straightforwardly
according to Eq.~(\ref{eq:conserved_quantities}).

%%%%%%%%%%%%%%%%%%%%%%%%%%%%%%%%%%%%%%%%%%%%%%%%%%%%%%%%%%%%%%%%%%%%%%%%%%%%%%%%%%%
%%%%%%%%%%%%%%%%%%%%%%%%%%%%%%%%%%%%%%%%%%%%%%%%%%%%%%%%%%%%%%%%%%%%%%%%%%%%%%%%%%%

\subsection{Pre-migration initial models}
\label{subsection:initial_models}

\begin{figure}
  \centerline{\includegraphics[width = 85 mm]{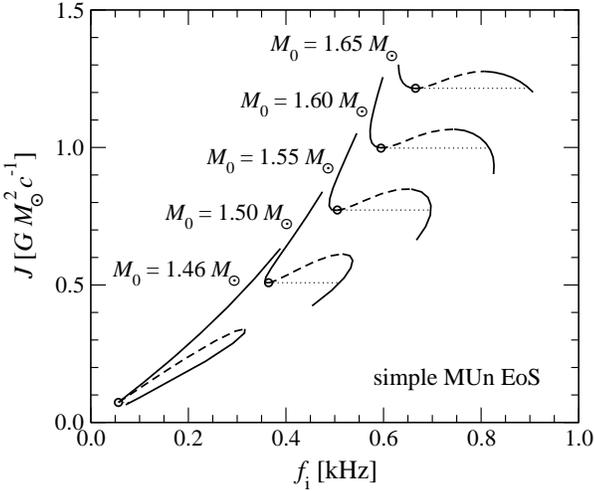}}
  \caption{Dependence of the total angular momentum $ J $ on the
    rotation frequency $ f_\mathrm{i} $ for uniformly rotating
    equilibrium models with the simple MUn EoS for various sequences
    of constant total baryon mass $ M_0 $. Solid/dashed lines specify
    the stable/unstable segments of the curve. The locations of the
    marginally stable initial models of the family US are marked by
    circles, and their transition path to the stable segment assuming
    conservation of $ J $ and uniform rotation, resulting in a change
    $ \Delta f $ in rotation frequency, is indicated by dotted lines.}
  \label{fig:back_bending_mun_eos}
\end{figure}

\begin{figure}
  \centerline{\includegraphics[width = 85 mm]{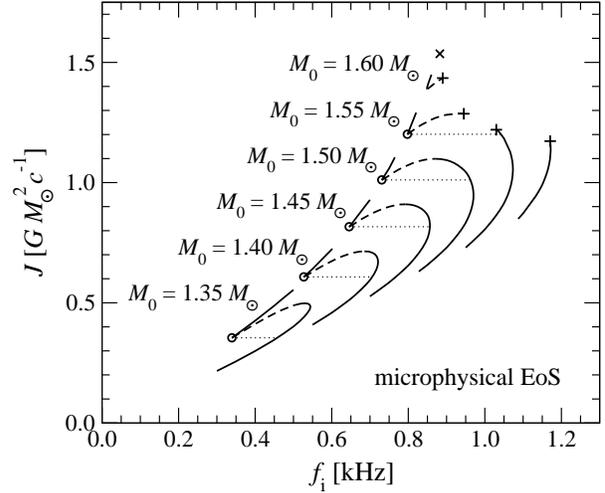}}
  \caption{Same as Fig.~\ref{fig:back_bending_mun_eos} but for models
    with the microphysical EoS. The locations of the marginally stable
    initial models of family UM are marked by circles. At the models
    marked by plus symbols, the Kepler limit interrupts the two
    sequences with $ M_0 = 1.55 \, M_\odot $ and
    $ M_0 = 1.60 \, M_\odot $. The transition path of the
    $ M_0 = 1.55 \, M_\odot $ sequence is not affected by this
    discontinuance, while for the $ M_0 = 1.60 \, M_\odot $ sequence
    there exists no model on the stable branch with the same $ J $ as
    the marginally stable one. Note that a fiducial line connecting
    all marginally stable models with this EoS would end with an
    $ M_0 = 1.639 \, M_\odot $ model rotating at the Kepler limit
    (marked by an x symbol).}
  \label{fig:back_bending_microphysical_eos}
\end{figure}

As shown by \citet{zdunik_06_a}, for the MUn EoS sequences of
uniformly rotating neutron star models with constant baryon mass $ M_0 $
exhibit a back bending phenomenon. As shown in
Fig.~\ref{fig:back_bending_mun_eos}, when plotting the total angular
momentum $ J $ against the rotation frequency $ f_\mathrm{i} $, for
each sequence there exists an unstable segment between a local minimum
and maximum of $ J $ (marked by dashed lines). As an isolated neutron
star spins down at constant $ M_0 $ along this evolutionary path, it
will migrate from the last marginally stable configuration (denoted by
a circle) to the location on the stable branch with the same total
angular momentum $ J $, provided that the migration proceeds
sufficiently rapid that $ J $ is conserved and that the migration does
not introduce a deviation of the rotation profile from uniformity.
While the first assumption is justified, as the migration is a fast
dynamic process, the condition of uniform rotation after the migration
has to be verified by numerical simulations. If the rotation profile
stays approximately uniform, then the migration path will proceed
closely along the dotted lines in
Fig.~\ref{fig:back_bending_mun_eos}. For the microphysical EoS, we
find the same back bending phenomenon for sequences of constant
$ M_0 $, as presented for a slightly different selection of values for
the gravitational mass in
Fig.~\ref{fig:back_bending_microphysical_eos}.

For our simulation of the dynamic migration due to back bending, we
construct initial models that correspond to the last marginally stable
configuration marked by circles\footnote{Even though
  Fig.~\ref{fig:back_bending_microphysical_eos} suggests that for
  models with the microphysical EoS, the local minimum of $ J $ for
  the marginally stable models coincides with the local minimum of
  $ f_\mathrm{i} $, a magnification of this plot reveals that this is
  not the case and the $ J $--$ f_\mathrm{i} $ curves actually are
  smooth everywhere.\label{footnote:back_bending_smoothness}} in
Figs.~\ref{fig:back_bending_mun_eos}
and~\ref{fig:back_bending_microphysical_eos}. The properties of these
initial models, which utilize the simple MUn EoS (family US) or the
microphysical EoS (family UM) are summarized in
Table~\ref{tab:unstable_initial_models}. Following the assumption that
during the spin down of isolated neutron stars viscous processes drive
the rotation profile towards uniformity, we initially assume rigid
rotation. Then each model is uniquely specified by its central energy
(or alternatively rest-mass) density and its rotation rate
$ T_\mathrm{i} / |W_\mathrm{i}| $, which is the ratio of rotational
energy to gravitational binding energy.

A remarkable property of the marginally stable initial models with the
microphysical EoS is that their central density is constant
irrespective of the baryon mass (see
Table~\ref{tab:unstable_initial_models}). This is a consequence of the
first-order phase transition in this EoS with a density jump between
$ \rho_1 $ and $ \rho_2 $. This transition destabilizes the neutron
star exactly when a (theoretically infinitesimally small) core of the
dense phase begins to form in its center\footnote{For numerical
  reasons, in our study the density discontinuity in the microphysical
  EoS is slightly softened by introducing a small pressure gradient,
  which leads to a threshold value
  $ \rho_\mathrm{c,i} = 9.8 \times 10^{14} \mathrm{\ g\ cm}^{-3} $
  that is a bit lower than $ \rho_2 $, where the dense phase
  begins. The smooth dependence of $ J $ on $ f_\mathrm{i} $ discussed
  in Footnote~\ref{footnote:back_bending_smoothness} is also a
  consequence of this pressure
  gradient.\label{footnote:pressure_gradient}}. In contrast, the MUn
EoS does not posses a first-order phase transition but rather a
continuous transition through a mixed phase without density jump.
Although the mixed phase also softens the EoS, this softening is not
as abrupt as in the microphysical EoS, but ``proportional'' to the
size of the soft core of the mixed phase. Thus the star loses
stability only if such a core of sizeable mass is already present in
the center, and the exact value of $ \rho_\mathrm{c} $ when the
marginally stable configuration is reached depends (albeit only
weakly) on the mass of the neutron star.

\begin{figure}
  \centerline{\includegraphics[width = 85 mm]{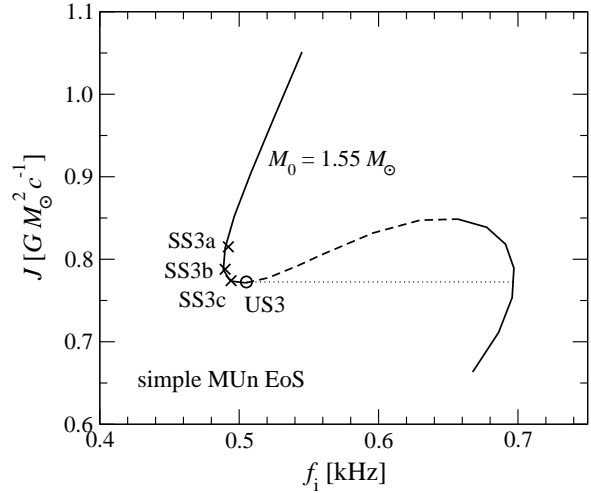}}
  \caption{Location of the three evolved stable models (marked by
    crosses) in the $ J $--$ f_\mathrm{i} $ diagram for uniformly
    rotating equilibrium models with the simple MUn EoS and
    $ M_0 = 1.55 \, M_\odot $ (see also
    Fig.~\ref{fig:back_bending_mun_eos}).}
  \label{fig:back_bending_mun_eos_stable}
\end{figure}

If the hydrodynamic and metric equations could be solved to
arbitrarily high precision, then the marginally stable models would
already migrate if subjected to an infinitely small perturbation. In
our numerical setup, however, truncation and round-off errors
together with the small errors from using the CFC approximation of the
exact metric equations introduce effects which make it hard to predict
if models in the close vicinity of a stable configuration are driven
over the edge of stability or not. In order to ensure that in our
numerical simulations the initial models undergo the migration
robustly in a finite evolution time, we choose to initially apply a
finite perturbation to the rest-mass density of the form
\begin{equation}
  \Delta \rho = a \, \sin \left( \frac{r}{r_\mathrm{atm}} \pi \right),
  \label{eq:perturbation}
\end{equation}
where $ a = 10^{-3} $ is the perturbation amplitude and
$ r_\mathrm{atm} $ is the (latitude dependent) boundary of the neutron
star to the surrounding low-density atmosphere. In order to assess the
evolutionary behavior of models near the stability limit, we augment
model US3 by the three models of family SS3 which are located close
to it still on the stable branch, as shown in
Fig.~\ref{fig:back_bending_mun_eos_stable}.

\begin{table*}
  \centering
  \caption{Summary of the set of migrating marginally stable initial
    models with the simple MUn EoS (family US) and the microphysical
    EoS (sequence UM) as well as the stable initial test models with
    the simple MUn EoS (family SS3). $ M_0 $ is the total baryon mass,
    $ M $ is the total gravitational mass, $ \mathcal{E}_\mathrm{c,i} $
    is the initial central energy density, $ \rho_\mathrm{c,i} $ is
    the initial central rest-mass density, $ f_\mathrm{i} $ is the
    initial rotation frequency, $ J $ is the total angular momentum,
    $ r_\mathrm{p,i} / r_\mathrm{e,i} $ is the initial ratio of polar
    radius to equatorial radius, and $ T_\mathrm{i} / |W_\mathrm{i}| $
    is the initial rotation rate.}
  \label{tab:unstable_initial_models}
  \begin{tabular}{@{}lcccccccccc@{}}
    \hline \\ [-1 em]
    Model &
    $ M_0 $ &
    $ M $ &
    $ \mathcal{E}_\mathrm{c,i} $ &
    $ \rho_\mathrm{c,i} $ &
    $ \rho_\mathrm{c,i} / \rho_1 $ &
    $ \rho_\mathrm{c,i} / \rho_2 $ &
    $ f_\mathrm{i} $ &
    $ J $ &
    $ r_\mathrm{p,i} / r_\mathrm{e,i} $ &
    $ T_\mathrm{i} / |W_\mathrm{i}| $ \\
    &
    [$ M_\odot $] &
    [$ M_\odot $] &
    [$ 10^{14} \mathrm{\ g\ cm}^{-3} $] &
    [$ 10^{14} \mathrm{\ g\ cm}^{-3} $] &
    &
    &
    [kHz] &
    [$ G M_\odot^2 c^{-1} $] &
    &
    [\%] \\ [0.2 em]
    \hline \\ [- 1 em]
    US1  & 1.46 & 1.35 &  10.22 & \z9.05 & 2.73 & 0.84 & 0.057 & 0.07 & 0.998 & 0.04 \\
    US2  & 1.50 & 1.39 &  10.39 & \z9.19 & 2.77 & 0.85 & 0.364 & 0.51 & 0.939 & 1.59 \\
    US3  & 1.55 & 1.44 &  10.39 & \z9.19 & 2.77 & 0.85 & 0.505 & 0.77 & 0.880 & 3.19 \\
    US4  & 1.60 & 1.48 &  10.22 & \z9.05 & 2.73 & 0.84 & 0.595 & 1.00 & 0.829 & 4.64 \\
    US5  & 1.65 & 1.53 &  10.27 & \z9.09 & 2.74 & 0.84 & 0.665 & 1.22 & 0.781 & 6.03 \\ [0.5 em]
    UM1  & 1.35 & 1.26 &  11.03 & \z9.80 & 2.95 & 0.91 & 0.340 & 0.36 & 0.953 & 1.19 \\
    UM2  & 1.40 & 1.30 &  11.03 & \z9.80 & 2.95 & 0.91 & 0.528 & 0.61 & 0.885 & 2.98 \\
    UM3  & 1.45 & 1.35 &  11.03 & \z9.80 & 2.95 & 0.91 & 0.646 & 0.82 & 0.823 & 4.63 \\
    UM4  & 1.50 & 1.39 &  11.03 & \z9.80 & 2.95 & 0.91 & 0.732 & 1.01 & 0.766 & 6.16 \\
    UM5  & 1.55 & 1.44 &  11.03 & \z9.80 & 2.95 & 0.91 & 0.798 & 1.21 & 0.711 & 7.58 \\
    UM6  & 1.60 & 1.48 &  11.03 & \z9.80 & 2.95 & 0.91 & 0.848 & 1.39 & 0.656 & 8.84 \\ [0.5 em]
    SS3a & 1.55 & 1.44 & \z8.21 & \z7.39 & 2.23 & 0.68 & 0.492 & 0.82 & 0.873 & 3.46 \\
    SS3b & 1.55 & 1.44 & \z8.87 & \z7.94 & 2.39 & 0.74 & 0.490 & 0.79 & 0.879 & 3.27 \\
    SS3c & 1.55 & 1.44 & \z9.60 & \z8.54 & 2.57 & 0.79 & 0.494 & 0.77 & 0.881 & 3.18 \\
    \hline
  \end{tabular}
\end{table*}

%%%%%%%%%%%%%%%%%%%%%%%%%%%%%%%%%%%%%%%%%%%%%%%%%%%%%%%%%%%%%%%%%%%%%%%%%%%%%%%%%%%
%%%%%%%%%%%%%%%%%%%%%%%%%%%%%%%%%%%%%%%%%%%%%%%%%%%%%%%%%%%%%%%%%%%%%%%%%%%%%%%%%%%
%%%%%%%%%%%%%%%%%%%%%%%%%%%%%%%%%%%%%%%%%%%%%%%%%%%%%%%%%%%%%%%%%%%%%%%%%%%%%%%%%%%

\section{Results}
\label{section:results}

%%%%%%%%%%%%%%%%%%%%%%%%%%%%%%%%%%%%%%%%%%%%%%%%%%%%%%%%%%%%%%%%%%%%%%%%%%%%%%%%%%%
%%%%%%%%%%%%%%%%%%%%%%%%%%%%%%%%%%%%%%%%%%%%%%%%%%%%%%%%%%%%%%%%%%%%%%%%%%%%%%%%%%%

\subsection{Dynamics of the migration process}
\label{subsection:migration_dynamics}

It is well known that the migration of a marginally stable nonrotating
spherical neutron star model with a polytropic EoS and adiabatic
coefficient $ \gamma = 2 $ leads to an expansion of the star
\citep[for a numerical simulation of this scenario, see
e.g.][]{font_02_a}. On the unstable branch, such a model possesses a
very compact structure with high central density $ \rho_\mathrm{c} $,
while the corresponding model with same baryon mass (and same
gravitational mass) on the stable branch has a much smaller value of
$ \rho_\mathrm{c} $. Consequently, the entire star significantly
expands in size and, after some ring-down pulsations following the
initial expansion, settles down to a less dense new equilibrium
configuration.

In stark contrast, in our case where two different EoSs are used and
the marginally stable initial neutron star models rotate rapidly, a
collapse instead of an expansion is expected. This follows directly
from the fact that during the migration the rotation frequency $ f $
increases as seen in Figs.~\ref{fig:back_bending_mun_eos}
and~\ref{fig:back_bending_microphysical_eos} while the total angular
momentum is conserved, if the uniform rotation profile is approximately
preserved throughout the evolution (which is a good approximation, as
later shown in Section~\ref{subsection:final_equilibrium_states}).
Still, as in the case of a nonrotating simple polytrope the new
equilibrium state -- now characterized by an increase in central
density compared to the initial value -- is overshot after the primary
contraction and is reached only after some ring-down pulsations.

\begin{figure}
  \centerline{\includegraphics[width = 85 mm]{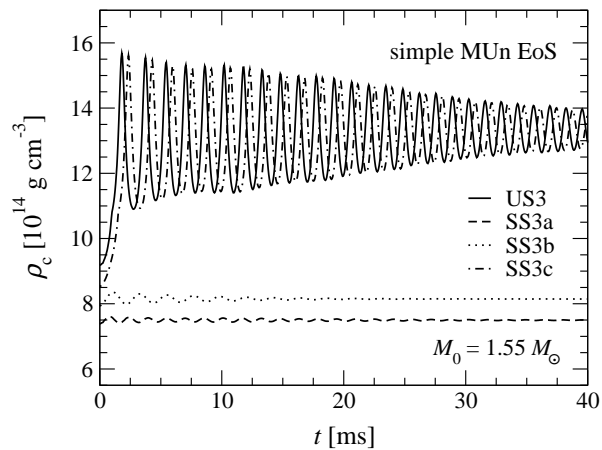}}
  \caption{Time evolution of the central density $ \rho_\mathrm{c} $
    for the marginally stable migration model US3 (solid line) and the
    three stable models SS3a, SS3b, and SS3c (dashed, dotted, and
    dash-dotted lines, respectively) with the simple MUn EoS and
    $ M_0 = 1.55 \, M_\odot $. Note that model SS3c actually migrates
    as well, resulting from the finite initial perturbation along with
    errors introduced by the numerical treatment and the CFC
    approximation.}
  \label{fig:density_evolution_mun_eos_stable}
\end{figure}

Before we present the collapse dynamics of the marginally stable
models that undergo migration, we first discuss the evolution of
models from the stable branch that are close to but still sufficiently
far from the stability limit, like models SS3a or SS3b. As expected
such models remain stable throughout the evolution and simply
oscillate around their equilibrium state due to the small perturbation
applied initially (see
Fig.~\ref{fig:density_evolution_mun_eos_stable}). However, the
amplitude of these damped pulsations already grows from model SS3a,
which is the one farthest from the marginally stable model US3, to
model SS3b that is closer to US3. Interestingly, model SS3c, which
lies very near model US3 but is still stable, already undergoes the
collapse to higher central densities and more rapid rotation that is
associated with the migration to the opposite stable segment. In this
case the errors from the numerical treatment and the CFC
approximation, and in particular the small but finite initial
perturbation suffice to push the neutron star model over the stability
limit. Its subsequent central density evolution is very similar to
that of the migrating marginally stable model US3, which is also shown
in Fig.~\ref{fig:density_evolution_mun_eos_stable}.

\begin{figure*}
  \centerline{\includegraphics[width = 180 mm]{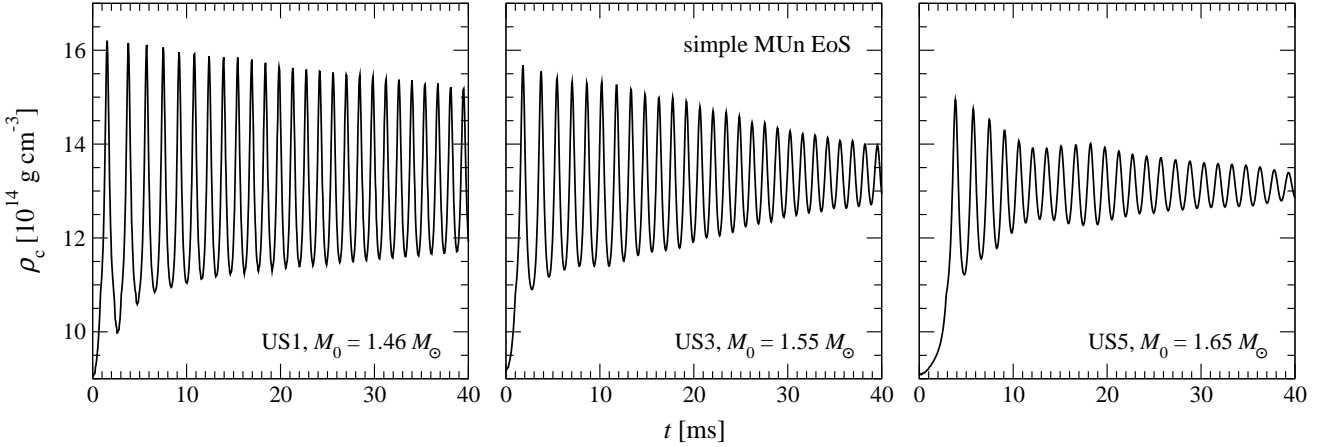}}
  \caption{Time evolution of the central density $ \rho_\mathrm{c} $
    for three migration models of family US with the simple MUn EoS.}
  \label{fig:density_evolution_mun_eos}
\end{figure*}

\begin{figure*}
  \centerline{\includegraphics[width = 180 mm]{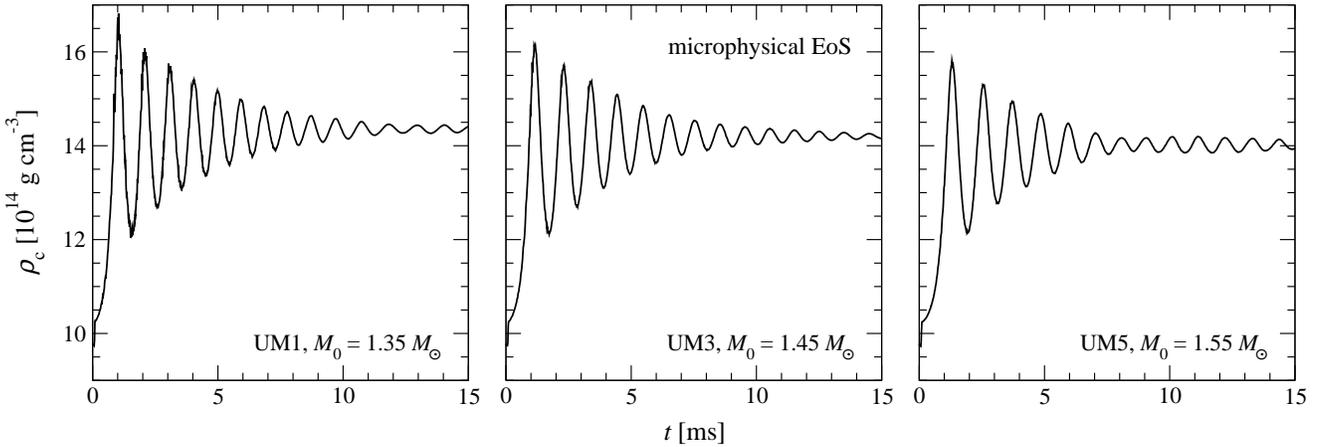}}
  \caption{Time evolution of the central density $ \rho_\mathrm{c} $
    for three migration models of family UM with the microphysical
    EoS.}
  \label{fig:density_evolution_microphysical_eos}
\end{figure*}

When comparing the time evolution of the central density for different
models of the same family, as presented in
Figs.~\ref{fig:density_evolution_mun_eos}
and~\ref{fig:density_evolution_microphysical_eos} for the two EoSs,
one recognizes a generic behavior. Within slightly more than
$ 1 \mathrm{\ ms} $, $ \rho_\mathrm{c} $ increases from its initial
value to a peak of $ \rho_\mathrm{c,p} \sim 15\mbox{\,--\,}16 \times
10^{14} \mathrm{\ g\ cm^{-3}} $, and subsequently undergoes damped
harmonic oscillations around an asymptotic value that lies between
$ \rho_\mathrm{c,p} \sim 13 $ and
$ 14 \times 10^{14} \mathrm{\ g\ cm^{-3}} $. The actual numbers for
the density increase in the peak and the final value are given in
Table~\ref{tab:migration_dynamics}.

\begin{table*}
  \centering
  \caption{Summary of the migration dynamics for models with the
    simple MUn EoS (family US) and the microphysical EoS (sequence
    UM). $ \rho_\mathrm{c,p} $ is the peak central density obtained at
    the first collapse peak, while $ \rho_\mathrm{c,f} $ is the
    asymptotic value for the central density after the pulsations have
    died out and $ \rho_\mathrm{c,sb} $ is the central density of the
    corresponding equilibrium model with the same value of $ J $ from
    the stable branch in Fig.~\ref{fig:back_bending_mun_eos}
    and~\ref{fig:back_bending_microphysical_eos}, respectively (for
    model UM6 that corresponding model is beyond the Kepler limit).
    $ M_{0,\rho > \rho_{1/2},\mathrm{i}} / M_0 $,
    $ M_{0,\rho > \rho_{1/2},\mathrm{p}} / M_0 $, and
    $ M_{0,\rho > \rho_{1/2},\mathrm{f}} / M_0 $ are fractions of
    baryon mass above the (EoS dependent) transition densities
    $ \rho_1 $ and $ \rho_2 $ at the initial time, the time of the
    maximum central density peak, and averaged at late evolution
    times, respectively. $ |h|_\mathrm{max} $ is the maximum
    gravitational wave strain, with the values without taking into
    account mode resonance effects  given in parentheses.}
  \label{tab:migration_dynamics}
  \begin{tabular}{@{}l@{~~}c@{~~}c@{\quad}c@{\quad}c@{\quad}c@{\quad}c@{\quad}c@{\quad}c@{\quad}c@{~~}c@{}}
    \hline \\ [-1 em]
%    Model &
%    $ \rho_\mathrm{c,p} $ &
%    $ \rho_\mathrm{c,p} / \rho_\mathrm{c,i} $ &
%    $ \rho_\mathrm{c,f} $ &
%    $ \rho_\mathrm{c,f,est} $ &
%    $ M_{0,\rho > \rho_1,\mathrm{i}} / M_0 $ &
%    $ M_{0,\rho > \rho_1,\mathrm{p}} / M_0 $ &
%    $ M_{0,\rho > \rho_1,\mathrm{f}} / M_0 $ &
%    $ M_{0,\rho > \rho_2,\mathrm{p}} / M_0 $ &
%    $ M_{0,\rho > \rho_2,\mathrm{f}} / M_0 $ &
%    $ |h|_\mathrm{max} $ \\ [0.2 em]
%    &
%    \highentry{[$ 10^{14} \mathrm{\ g\ cm}^{-3} $]} &
%    &
%    \highentry{[$ 10^{14} \mathrm{\ g\ cm}^{-3} $]} &
%    \highentry{[$ 10^{14} \mathrm{\ g\ cm}^{-3} $]} &
%    &
%    &
%    &
%    &
%    &
%    $ \displaystyle \left[ \!\!\!
%      \begin{array}{c}
%        10^{-21} \\ [-0.2 em]
%        \mathrm{\ at\ 10\ kpc}
%      \end{array}
%    \! \right] $ \\ [0.8 em]
    Model &
    \multicolumn{1}{p{1.5 cm}}{\centering $ \rho_\mathrm{c,p} $ \\ [0.5 em] $ \displaystyle \left[ 10^{14} \frac{\mathrm{g}}{\mathrm{cm}^{3}} \right] $} &
    \lowentry{$ \displaystyle \frac{\rho_\mathrm{c,p}}{\rho_\mathrm{c,i}} $} &
    \multicolumn{1}{p{1.5 cm}}{\centering $ \rho_\mathrm{c,f} $ \\ [0.5 em] $ \displaystyle \left[ 10^{14} \frac{\mathrm{g}}{\mathrm{cm}^{3}} \right] $} &
    \multicolumn{1}{p{1.5 cm}}{\centering $ \rho_\mathrm{c,sb} $ \\ [0.5 em] $ \displaystyle \left[ 10^{14} \frac{\mathrm{g}}{\mathrm{cm}^{3}} \right] $} &
    \lowentry{$ \displaystyle \frac{M_{0,\rho > \rho_1,\mathrm{i}}}{M_0} $} &
    \lowentry{$ \displaystyle \frac{M_{0,\rho > \rho_1,\mathrm{p}}}{M_0} $} &
    \lowentry{$ \displaystyle \frac{M_{0,\rho > \rho_1,\mathrm{f}}}{M_0} $} &
    \lowentry{$ \displaystyle \frac{M_{0,\rho > \rho_2,\mathrm{p}}}{M_0} $} &
    \lowentry{$ \displaystyle \frac{M_{0,\rho > \rho_2,\mathrm{f}}}{M_0} $} &
    $ |h|_\mathrm{max} $ \\ [2.7 em]
    \hline \\ [- 1 em]
    US1 & \z16.24 & 1.796 & \zz13.48 & \zz13.69 (2\%) & 0.35 & 0.71 & 0.58 & 0.35 & 0.17 & 0.48 (0.12) \\
    US2 & \z15.99 & 1.741 & \zz13.33 & \zz13.62 (2\%) & 0.35 & 0.70 & 0.58 & 0.33 & 0.17 & 3.69 (2.81) \\
    US3 & \z15.69 & 1.708 & \zz13.33 & \zz13.55 (2\%) & 0.35 & 0.68 & 0.57 & 0.31 & 0.17 & 7.97 (3.94) \\
    US4 & \z15.33 & 1.725 & \zz13.48 & \zz13.50 (1\%) & 0.33 & 0.66 & 0.56 & 0.29 & 0.16 & 4.35  \zzzp \\
    US5 & \z14.94 & 1.644 & \zz13.09 & \zz13.45 (3\%) & 0.34 & 0.64 & 0.54 & 0.26 & 0.15 & 4.62  \zzzp \\ [0.5 em]
    UM1 & \z16.84 & 1.718 & \zz14.31 & \zz14.28 (1\%) & 0.01 & 0.46 & 0.32 & 0.43 & 0.28 & 4.50 (1.42) \\
    UM2 & \z16.37 & 1.670 & \zz14.11 & \zz14.28 (1\%) & 0.01 & 0.45 & 0.31 & 0.41 & 0.28 & 6.90 (4.25) \\
    UM3 & \z16.16 & 1.649 & \zz14.11 & \zz14.27 (1\%) & 0.01 & 0.43 & 0.30 & 0.40 & 0.27 & 6.69 (5.90) \\
    UM4 & \z16.00 & 1.633 & \zz14.11 & \zz14.27 (1\%) & 0.01 & 0.42 & 0.29 & 0.38 & 0.26 & 6.50  \zzzp \\
    UM5 & \z15.83 & 1.615 & \zz13.92 & \zz14.29 (3\%) & 0.01 & 0.40 & 0.28 & 0.36 & 0.25 & 6.89  \zzzp \\
    UM6 & \z15.61 & 1.597 & \zz13.92 &            --- & 0.01 & 0.38 & 0.27 & 0.35 & 0.24 & 7.57  \zzzp \\
    \hline
  \end{tabular}
\end{table*}

The baryon mass fractions in Table~\ref{tab:migration_dynamics} also
demonstrate that initially no fluid element in the neutron star has a
density above $ \rho_2 $, which is consistent with the values for
$ \rho_\mathrm{c,i} $ given in
Table~\ref{tab:unstable_initial_models}, while around $ 35\% $ and
$ 60\% $ of the stellar mass is at a density above $ \rho_1 $ for the
MUn EoS and the microphysical EoS, respectively. Later in the
evolution, the mass of the star above $ \rho_1 $ grows noticeably, and
even a significant fraction of $ M_0 $ has a density greater than
$ \rho_2 $.

When comparing Figs.~\ref{fig:density_evolution_mun_eos}
and~\ref{fig:density_evolution_microphysical_eos} it is apparent that
while the density oscillations of models with the MUn EoS induced by the
migration are somewhat damped, the damping in the models with the
microphysical EoS is much stronger\footnote{Note the different time
  scales in the two figures!}. After only a few cycles, the pulsation
have died out and $ \rho_\mathrm{c} $ (as well as other hydrodynamic
and metric quantities at various locations in the star) remain
practically constant. By performing simulations with significantly
higher grid resolution, we have ensured that this effect is not caused
by insufficient numerical accuracy. In addition, since our code
utilizes Riemann flux solvers, which can handle discontinuities in the
hydrodynamic quantities without problem \cite[see e.g.\ the
relativistic shock tube test presented in][]{dimmelmeier_02_a}, we can
exclude that the pressure jump in the microphysical EoS leads to
numerically questionable results. By using a relative convergence
criterion of $ 10^{-12} $, we also ascertain that the recovery routine
for the primitive quantities is performed to sufficient precision.

When investigating the strong damping of post-migration pulsations
in the models of family UM with the microphysical EoS, we find strong
evidence that it is a result of a very efficient physical damping
mechanism that is reflected in our code. As shown in detail in
Appendix~\ref{appendix:pulsation_damping}, the kinetic energy stored
in the pulsations is dissipated during the compression phase of each
pulsation, as matter passes through the density jump from $ \rho_1 $
to $ \rho_2 $ associated to the first-order phase transition in this
EoS. While in reality matter flows into the dense core at supersonic
speed and is heated up by shock dissipation, in our
simulations the flow remains subsonic. Still, the large density jump
results in a rather steep radial inflow velocity profile, leading to
numerical dissipation of comparable strength. If the density jump in
the EoS is gradually removed, dissipation continuously decreases to
the normal level expected for a code based on HRSC methods.

Furthermore, non-equilibrium processes produce entropy via
heating. This can be represented by bulk viscosity, but as the
corresponding additional dissipation is much smaller than the
dissipation caused by the phase transition (see
Appendix~\ref{subappendix:bulk_viscosity_damping}), it is admissible
to neglect this mechanism as it is done in our simulations. Another,
less important mechanism that is responsible for the damping of
post-migration pulsations is the continuous but weak conversion of
kinetic energy stored in the pulsations into internal energy by
numerical dissipation in the bulk of the star (excluding the parts of
the star that are affected by a possible density jump in the
EoS). However, as this process is equally active in all models
irrespective of the EoS, in the models with the microphysical EoS it
can at most explain damping in a small extent similar to what is
observed in the models with the simple MUn EoS, which have a much
larger damping time scale. In contrast to some of the rapidly rotating
neutron star models discussed by \citet{abdikamalov_09_a}, which
undergo a phase-transition-induced collapse and whose post-collapse
pulsations are mainly damped by the shedding of mass from the surface
to the atmosphere, we do not find any evidence for the effect of
damping by  shedding in any of our simulations. The loss of mass
across the boundary of our neutron star models remains well below
$ 1\% $ until we deliberately terminate the evolution at
$ t_\mathrm{f} = 50 \mathrm{\ ms} $.

It is remarkable that according to Table~\ref{tab:migration_dynamics}
all respective models of family US and UM show a generic ``radial''
collapse behavior, expressed by a very similar relative increase in
central density and fractions of baryon mass above the respective (EoS
dependent) transition densities at the collapse peak and the final
equilibrium, although the change $ \Delta f $ in rotation frequency
along the migration path in the $ J $--$ f_\mathrm{i} $ diagram is
rather different (see Figs.~\ref{fig:back_bending_mun_eos}
and~\ref{fig:back_bending_microphysical_eos}). For the two
investigated EoSs, the balance between pressure, gravitational, and
centrifugal forces already lead to almost identical values of these
mass fractions and the central density\footnote{In the case of the
  microphysical EoS, $ \rho_\mathrm{c,i} $ is indeed independent of
  $ M_0 $ as explained in Section~\ref{subsection:initial_models}.} in
the marginally stable initial models (see also
Table~\ref{tab:unstable_initial_models}), and apparently this does not
change during the collapse. We also point out that while the migration
lets $ \rho_\mathrm{c} $ increase by about $ 60\mbox{\,--\,}70\% $ to
its peak value and $ 40\mbox{\,--\,}50\% $ to the final equilibrium
value, the \emph{relative} changes in the rotation rate
$ T_\mathrm{i} / |W_\mathrm{i}| $ are comparatively small with values
of about $ 9\mbox{\,--\,}12\% $ and $ 6\mbox{\,--\,}7\% $,
respectively.

%%%%%%%%%%%%%%%%%%%%%%%%%%%%%%%%%%%%%%%%%%%%%%%%%%%%%%%%%%%%%%%%%%%%%%%%%%%%%%%%%%%
%%%%%%%%%%%%%%%%%%%%%%%%%%%%%%%%%%%%%%%%%%%%%%%%%%%%%%%%%%%%%%%%%%%%%%%%%%%%%%%%%%%

\subsection{Final equilibrium states after the migration and spectrum
  of the pulsations}
\label{subsection:final_equilibrium_states}

If the uniformity of the profile rotation is maintained during the
dynamic migration and thus the collapsed neutron star in its new final
equilibrium state also rotates rigidly, as already speculated by
\citet{zdunik_06_a}, then the migration essentially follows the
path marked by the horizontal dotted lines in
Figs.~\ref{fig:back_bending_mun_eos}
and~\ref{fig:back_bending_microphysical_eos}, modified only negligibly
by a tiny change in $ J $ due to gravitational wave emission (which
however cannot be modeled with our formulation of the metric
equations).

\begin{figure*}
  \centerline{\includegraphics[width = 180 mm]{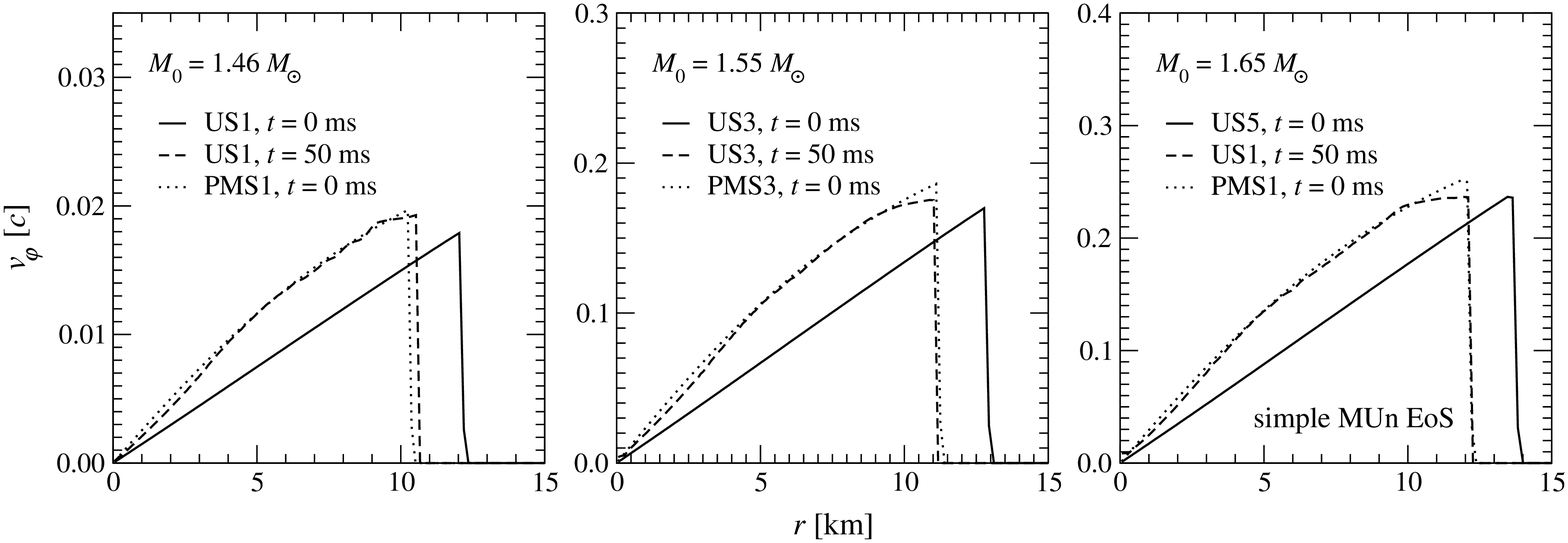}}
  \caption{Radial profiles in the equatorial plane of the rotation
    velocity $ v_\phi $ for three models with the simple MUn EoS. The
    solid and dashed lines are at the initial and final evolution
    time, respectively, for migration models of family US, while the
    dotted lines show $ v_\phi $ at the initial evolution time for
    stable post-migration models of family PMS.}
  \label{fig:rotation_velocity_profile_mun_eos}
\end{figure*}

\begin{figure*}
  \centerline{\includegraphics[width = 180 mm]{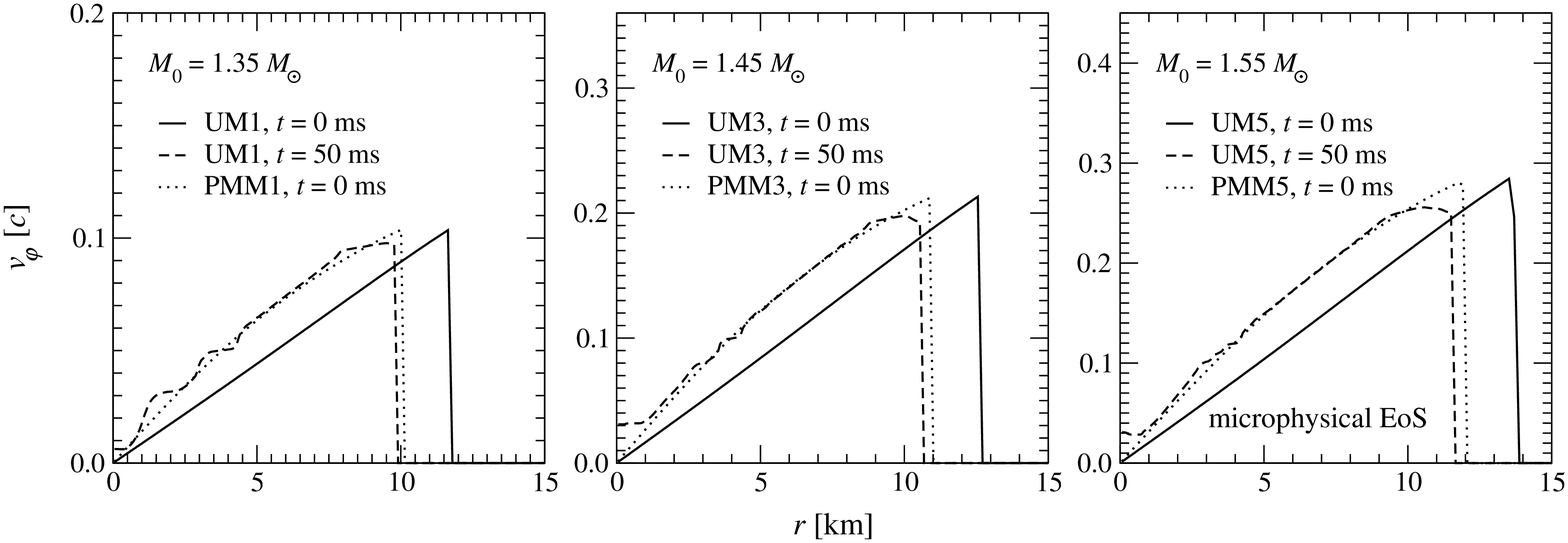}}
  \caption{Radial profiles in the equatorial plane of the rotation
    velocity $ v_\phi $ for three models with the microphysical EoS.
    The solid and dashed lines are at the initial and final evolution
    time, respectively, for migration models of family UM, while the
    dotted lines show $ v_\phi $ at the initial evolution time for
    stable post-migration models of family PMM.}
  \label{fig:rotation_velocity_profile_microphysical_eos}
\end{figure*}

With our fully dynamic code we are now able to test this assumption of
preservation of the rotation profile during the collapse and the
subsequent ring-down. In
Figs.~\ref{fig:rotation_velocity_profile_mun_eos}
and~\ref{fig:rotation_velocity_profile_microphysical_eos} the radial
profiles of the rotation velocity $ v_\varphi $ in the equatorial
plane are plotted both for the initial model (where they obviously
scale linearly with radius $ r $) and at the end of the evolution at
$ 50 \mathrm{\ ms} $ for the simple MUn EoS and the microphysical
EoS, respectively. In both cases the degree of differential rotation
created by the migration is rather small, although the central density
increases by up to 50\%. It is actually possible to approximate the
final rotation profile rather accurately by an analytic law for the
specific angular momentum $ j $ \citep[see e.g.][]{komatsu_89_a},
\begin{equation}
  j = A^2 (\Omega_\mathrm{c} - \Omega),
  \label{eq:differential_rotation_law}
\end{equation}
where $ A $ parametrizes the degree of differential rotation (stronger
differentiality with decreasing $ A $) and $ \Omega_\mathrm{c} $ is
the value of the angular velocity $ \Omega $ at the center. This
simple relation is widely used in general relativistic hydrodynamics
to model differential rotation in compact stars. The differential
rotation length scale $ A $ can be put into relation with the
equatorial radius $ r_\mathrm{e} $ in isotropic coordinates by
defining the dimensionless quantity $ \hat{A} $. If $ \hat{A} \ll 1 $,
then the rotation profile is very differential (with a larger specific
angular momentum distribution at small distances to the rotation
axis), while $ \hat{A} \gg 1 $ leads towards the limit of uniform
rotation. For the approximated post-migration rotation curves of our
models we find values $ \hat{A} $ that are always larger than $ 1 $
and reach almost $ 3 $ for some models (as presented in
Table~\ref{tab:post_migration_initial_models}) which confirms the
conclusions from the visual analysis of
Figs.~\ref{fig:rotation_velocity_profile_mun_eos}
and~\ref{fig:rotation_velocity_profile_microphysical_eos}. Despite the
creation of some differential rotation during the dynamic migration,
the post-migration models are thus rather well approximated by the
equilibrium models on the stable branch in the $ J $--$ f_\mathrm{i} $
diagram (given by the right end points of the dotted lines in
Figs.~\ref{fig:back_bending_mun_eos}
and~\ref{fig:back_bending_microphysical_eos}) that have the same value
of $ J $ as the marginally stable initial model. Further evidence for
this conclusion is given by the observation that the final central
density $ \rho_\mathrm{c,f} $ of any post-migration model differs
from the central density $ \rho_\mathrm{c,sb} $ of the corresponding
model on the stable branch by at most 3\% (see
Table~\ref{tab:migration_dynamics}).

Although the pulsations triggered by the initial collapse are only
slowly damped for the models with the simple MUn EoS, they will
eventually die out, and a stable equilibrium will be reached. With the
knowledge of an approximate analytic solution for the final rotation
profile we are thus able to compute stable equilibrium models which
should closely match the post-migration configurations obtained from
the dynamic evolution. For this we utilize the same method employed to
construct the marginally stable initial models for the migration
simulations, with the important difference that now slightly
differential rotation profiles with the values for $ \hat{A} $ from
Table~\ref{tab:post_migration_initial_models} are needed. Using the
same EoS as for the respective migration model, only one additional
parameter to uniquely determine each of these new equilibrium models
is left, like e.g.\ the baryon mass $ M_0 $, the gravitational mass
$ M $, the total angular momentum $ J $, or the central (energy or
rest-mass) density $ \mathcal{E}_\mathrm{c} $ or $ \rho_\mathrm{c} $.

\begin{table*}
  \centering
  \caption{Summary of the set of stable post-migration initial models
    with the simple MUn EoS (family PMS) and the microphysical EoS
    (family PMM). $ f_\mathrm{c,i} $ is the central rotation
    frequency, $ \hat{A} $ is the differential rotation length scale,
    and the other quantities are explained in
    Table~\ref{tab:unstable_initial_models}. Note that as required,
    the quantities $ M_0 $, $ M $ and $ J $ are almost identical to
    the respective marginally stable models.}
  \label{tab:post_migration_initial_models}
  \begin{tabular}{@{}lccccccccc@{}}
    \hline \\ [-1 em]
    Model &
    $ M_0 $ &
    $ M $ &
    $ \mathcal{E}_\mathrm{c,i} $ &
    $ \rho_\mathrm{c,i} $ &
    $ f_\mathrm{c,i} $ &
    $ J $ &
    $ \hat{A} $ &
    $ r_\mathrm{p,i} / r_\mathrm{e,i} $ &
    $ T_\mathrm{i} / |W_\mathrm{i}| $ \\
    &
    [$ M_\odot $] &
    [$ M_\odot $] &
    [$ 10^{14} \mathrm{\ g\ cm}^{-3} $] &
    [$ 10^{14} \mathrm{\ g\ cm}^{-3} $] &
    [kHz] &
    [$ G M_\odot^2 c^{-1} $] &
    &
    &
    [\%] \\ [0.2 em]
    \hline \\ [- 1 em]
    PMS1 & 1.45 & 1.35 & 15.93 & 13.53 & 0.089 & 0.07 & 2.7 & 0.998 & 0.04 \\
    PMS2 & 1.50 & 1.39 & 15.81 & 13.44 & 0.549 & 0.50 & 2.8 & 0.930 & 1.66 \\
    PMS3 & 1.56 & 1.44 & 15.67 & 13.33 & 0.810 & 0.81 & 2.5 & 0.850 & 3.72 \\
    PMS4 & 1.59 & 1.47 & 15.53 & 13.23 & 0.882 & 0.97 & 2.7 & 0.810 & 4.77 \\
    PMS5 & 1.66 & 1.54 & 15.34 & 13.09 & 1.025 & 1.27 & 2.5 & 0.740 & 6.76 \\ [0.5 em]
    PMM1 & 1.35 & 1.26 & 16.73 & 14.34 & 0.498 & 0.34 & 2.5 & 0.950 & 1.18 \\
    PMM2 & 1.40 & 1.30 & 16.60 & 14.23 & 0.763 & 0.58 & 2.5 & 0.880 & 2.91 \\
    PMM3 & 1.45 & 1.34 & 16.51 & 14.18 & 0.945 & 0.80 & 2.5 & 0.810 & 4.73 \\
    PMM4 & 1.50 & 1.39 & 16.38 & 14.08 & 1.063 & 0.99 & 2.5 & 0.750 & 6.34 \\
    PMM5 & 1.54 & 1.43 & 16.30 & 14.01 & 1.118 & 1.15 & 2.7 & 0.700 & 7.60 \\
    PMM6 & 1.59 & 1.47 & 16.11 & 13.87 & 1.195 & 1.34 & 2.5 & 0.650 & 9.01 \\
    \hline
  \end{tabular}
\end{table*}

For practical reasons we chose $ \rho_\mathrm{c} $ as the
parameter of choice, which we select such that it equals the value of
$ \rho_\mathrm{c,f} $ in Table~\ref{tab:migration_dynamics}. With this
method for each migration model of the two families US and UM an
accompanying equilibrium model that resembles the final state is
constructed. We label the resulting families of approximative
post-merger models by PMS and PMM for the simple MUn EoS and the
microphysical EoS, respectively. The parameters for these models are
summarized in Table~\ref{tab:post_migration_initial_models}.
The excellent agreement in both $ M_0 $ and $ M $ as well as $ J $
between the respective members of families of marginally stable
initial models and post-migration equilibrium models (compare the
values given in Tables~\ref{tab:unstable_initial_models}
and~\ref{tab:post_migration_initial_models}) illustrates the validity
of this approach\footnote{We note that it makes no sense to locate
  these post-migration equilibrium models in the
  $ J $--$ f_\mathrm{i} $ diagram in
  Figs.~\ref{fig:back_bending_mun_eos}
  and~\ref{fig:back_bending_microphysical_eos}, as they possess a
  small but non-negligible amount of differential rotation.}.
Therefore, it is possible to very accurately describe the neutron star
at the end point of the migration by a simple equilibrium model
obeying an analytic rotation law, which facilitates a variety of ways
to further analyze the outcome of the migration process.

This method even works for the peculiar model UM6 with the
microphysical EoS, whose marginally stable initial configuration has
no counterpart on the stable branch at constant $ J $ and respectively
higher $ f $ in the $ J $--$ f_\mathrm{i} $ diagram for uniformly
rotating equilibrium models in
Fig.~\ref{fig:back_bending_microphysical_eos}, as models in this
segment of the sequence are beyond the Kepler limit. It is noteworthy
that model UM5 exhibits the same migration dynamics as the other
models of its family, without contracting to excessively high
densities (or even collapsing a black hole) or being subject to mass
shedding. The small (albeit nonzero) development of a differential
rotation profile during the migration suffices to create a stable
final equilibrium state that is located within the Kepler limit. We
are thus able to construct the corresponding post-migration
equilibrium model PMM6, whose value $ \hat{A} = 2.5 $ for the
differential rotation parameter is typical for family UM (see
Table~\ref{tab:post_migration_initial_models}).

We find that even the extremal marginally stable model with
$ M_0 = 1.639 \m M_\odot $, which already initially rotates at the
Kepler limit (see Fig.~\ref{fig:back_bending_microphysical_eos})
migrates as the other models to a stable, differentially rotating
equilibrium configuration without mass shedding. The segments of the $
M_0 = \mathrm{const.} $ sequences in the $ J $--$ f_\mathrm{i} $
diagram that are beyond the Kepler limit are therefore irrelevant for
the actual migration process, as the fiducial migration path indicated
by dotted lines in that figure requires exact preservation of uniform
rotation throughout the migration, which is not obeyed during a fully
dynamic evolution. We note that the marginally stable models with the
simple MUn EoS also end at the Kepler limit with a model that,
however, has a rather large baryon mass $ M_0 = 1.899 M_\odot $. As
this model's total angular momentum $ J = 2.30 $ is very large as
well, we refrain from including it in the respective
$ J $--$ f_\mathrm{i} $ diagram in
Fig.~\ref{fig:back_bending_mun_eos}.

%%%%%%%%%%%%%%%%%%%%%%%%%%%%%%%%%%%%%%%%%%%%%%%%%%%%%%%%%%%%%%%%%%%%%%%%%%%%%%%%%%%
%%%%%%%%%%%%%%%%%%%%%%%%%%%%%%%%%%%%%%%%%%%%%%%%%%%%%%%%%%%%%%%%%%%%%%%%%%%%%%%%%%%

\subsection{Analysis of the pulsations modes}
\label{subsection:mode_analysis}

\begin{figure*}
  \centerline{\includegraphics[width = 180 mm]{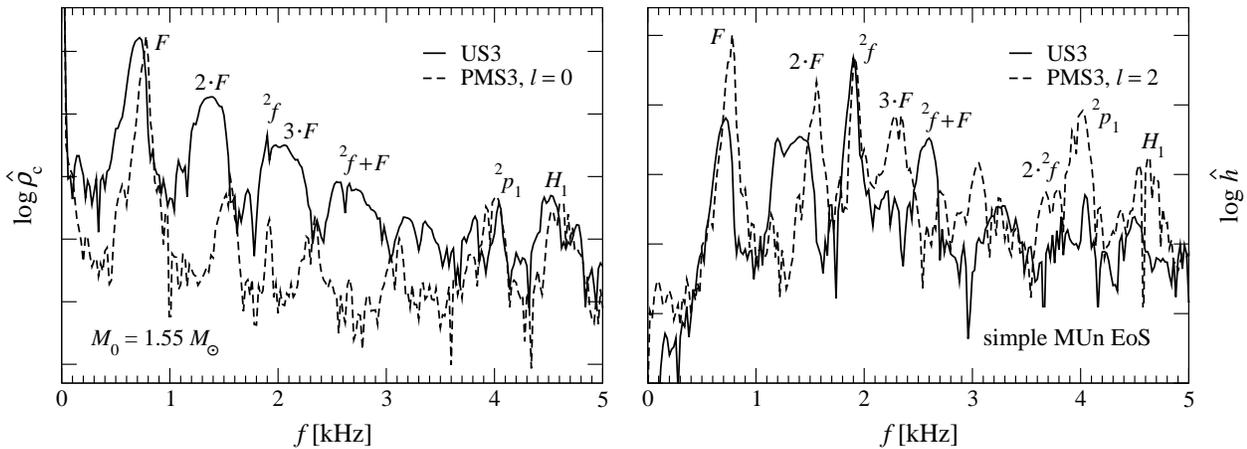}}
  \caption{Logarithmic frequency spectrum (i.e.\ Fourier transform as
    indicated by the hat symbol) of the time evolution of the central
    density $ \rho_\mathrm{c} $ (left panel) and the gravitational
    wave strain $ h $ (right panel) for the migration model US3 (solid
    line) and the associated post-migration equilibrium model PMS3
    (dashed line, excited with an $ l = 0/2 $ trial eigenfunction for
    $ \rho_\mathrm{c}/h $, respectively) with the simple MUn EoS. The
    vertical scale is one order of magnitude per tick mark. Several
    prominent quasi-normal modes and their nonlinear couplings are
    marked. The data have been scaled such that the peaks of the
    $ F $-mode (left panel) and the $ {}^{2\!}f $-mode (right panel)
    have the same height for the two models US3 and PMS3.}
  \label{fig:spectrum_mun_eos}
\end{figure*}

A clear example of the benefit of being able to approximate the final
post-migration states of our models by simple equilibrium
configurations is the possibility to comprehensively examine the
quasi-normal pulsation modes triggered by the migration by perturbing
those respective post-migration equilibrium models. It is difficult in
some cases to unambiguously identify the mode nature and its
eigenfunction structure directly from the dynamic simulation of the
migration itself, as both the absolute and relative excitation of
the various modes is not under control, and the pulsation spectrum is
rather complex (see the solid lines in
Fig.~\ref{fig:spectrum_mun_eos}). In contrast, in the corresponding
equilibrium models, by applying a initially a selection of small trial
eigenfunctions of defined parity, we can excite a range of particular
quasi-normal modes in the linear regime with controlled amplitude. For
$ l = 0 $ oscillations, we set the radial velocity to
\begin{equation}
  v_r = a \, \sin \left( \frac{r}{r_\mathrm{atm}} \pi \right)
  \label{eq:trial_eigenfunction_0}
\end{equation}
with amplitude $ a = -0.005 $, and for $ l = 2 $ we set the angular
velocity to
\begin{equation}
  v_\theta = a \, \sin \theta \, \cos \theta \,
  \sin \left( \frac{r}{r_\mathrm{atm}} \pi \right)
  \label{eq:trial_eigenfunction_2}
\end{equation}
with $ a = 0.01 $. In order to improve the results obtained from these
perturbed equilibrium models even more, in difficult
cases\footnote{For rapid rotation, the quasi-radial modes acquire a
  strong quadrupolar contribution, making an unambiguous identification
  (particularly of the first harmonic mode $ H_1 $) a demanding task.}
we apply a further recycling run for each of the main quasi-normal
modes, which are the fundamental quasi-radial $ F $-mode, the
fundamental quadrupolar $ {}^{2\!}f $-mode, and their respective
harmonics, the $ H_1 $-mode and the $ {}^{2\!}p_1 $-mode \cite[for
this we essentially follow the technique described
in][]{dimmelmeier_06_a}. This method facilitates an accurate
extraction of both the frequency and the eigenfunction of each mode,
and thus allows us to clearly identify the mode nature. The
improvement in the clarity of the pulsation spectrum from using the
trial eigenfunctions alone is apparent from the dashed lines in
Fig.~\ref{fig:spectrum_mun_eos}, where $ l = 0/2 $ parity was used in
the left/right panel, respectively, as quasi-radial modes are
predominantly visible in quantities like the density, while
quadrupolar modes are prominent for instance in the waveform.

\begin{table*}
  \centering
  \caption{Frequencies for quasi-normal modes and some of their
    nonlinear self-couplings for the migration models of the family US
    and the post-migration equilibrium models of the family PMS with
    the simple MUn EoS. $ f_F $, $ f_{{}^{2\!}f} $, $ f_{H_1} $, and
    $ f_{{}^{2\!}p_1} $ are the frequencies of the fundamental
    quasi-radial mode, the fundamental quadrupolar mode, and their
    respective first harmonics, while the other modes are nonlinear
    self-couplings. For the quasi-normal modes the relative
    differences in frequency between the model of the US and PMS
    family are also given. For some models, particular modes could not
    be extracted with sufficient confidence, probably due to
    interaction with nearby other modes.}
  \label{tab:frequencies_mun_eos}
  \begin{tabular}{@{}lc@{\quad}rc@{\quad}rc@{\quad}rc@{\quad}rccc@{}}
    \hline \\ [-1 em]
    Model &
    \multicolumn{2}{c}{$ f_F $} &
    \multicolumn{2}{c}{$ f_{{}^{2\!}f} $} &
    \multicolumn{2}{c}{$ f_{H_1} $} &
    \multicolumn{2}{c}{$ f_{{}^{2\!}p_1} $} &
    $ f_{2 \cdot\! F} $ &
    $ f_{3 \cdot\! F} $ &
    $ f_{2 \cdot {}^{2\!}f} $ \\
    &
    \multicolumn{2}{c}{[kHz]} &
    \multicolumn{2}{c}{[kHz]} &
    \multicolumn{2}{c}{[kHz]} &
    \multicolumn{2}{c}{[kHz]} &
    [kHz] &
    [kHz] &
    [kHz] \\ [0.2 em]
    \hline \\ [- 1 em]
    US1  &
    0.713 & \lowentry{19\%} &
    1.817 &  \lowentry{4\%} &
    4.792 &  \lowentry{3\%} &
    4.173 &  \lowentry{1\%} &
    1.407 &
    2.107 &
    3.625
    \\
    PMS1 &
    0.848 &                &
    1.893 &                &
    4.901 &                &
    4.231 &                &
    1.705 &
    2.554 &
    3.795
    \\ [0.5 em]
    US2  &
    0.723 & \lowentry{10\%} &
    1.867 &  \lowentry{2\%} &
      --- &  \lowentry{---} &
    4.320 &  \lowentry{2\%} &
    1.429 &
    2.140 &
    3.665
    \\
    PMS2 &
    0.794 &                 &
    1.907 &                 &
    4.748 &                 &
    4.166 &                 &
    1.609 &
    2.412 &
    3.824
    \\ [0.5 em]
    US3  &
    0.728 &  \lowentry{7\%} &
    1.909 &  \lowentry{1\%} &
    4.483 &  \lowentry{3\%} &
    4.048 &  \lowentry{1\%} &
    1.389 &
    2.070 &
    3.817
    \\
    PMS3 &
    0.777 &                 &
    1.924 &                 &
    4.619 &                 &
    4.024 &                 &
    1.562 &
    2.350 &
    3.870
    \\ [0.5 em]
    US4  &
    0.716 &  \lowentry{2\%} &
    1.926 &  \lowentry{1\%} &
    4.499 &  \lowentry{1\%} &
    3.875 &  \lowentry{5\%} &
    1.421 &
    2.131 &
      ---
    \\
    PMS4 &
    0.730 &                 &
    1.914 &                 &
    4.536 &                 &
    3.893 &                 &
    1.456 &
    2.192 &
      ---
    \\ [0.5 em]
    US5  &
    0.663 &  \lowentry{7\%} &
    1.922 &  \lowentry{2\%} &
    4.517 &  \lowentry{1\%} &
    3.673 &  \lowentry{1\%} &
    1.318 &
    1.974 &
      ---
    \\
    PMS5 &
    0.617 &                 &
    1.890 &                 &
    4.464 &                 &
    3.635 &                 &
    1.238 &
    1.877 &
      ---
    \\
    \hline
  \end{tabular}
\end{table*}

The frequencies of the four main quasi-normal modes (the fundamental
$ F $-mode and $ {}^{2\!}f $-mode, and the harmonic $ H_1 $-mode and
$ {}^{2\!}p_1 $-mode) extracted from the migration models and the
respective values from their respective recreated post-migration
equilibrium configurations excited by small amplitude perturbations
along with the major nonlinear self-couplings are presented in
Table~\ref{tab:frequencies_mun_eos} for the simple MUn Eos and in
Table~\ref{tab:frequencies_microphysical_eos} for the microphysical
Eos. Except for the fundamental quasi-radial $ F $-mode, which
nevertheless shows satisfactory agreement, the equilibrium models
reproduce the frequencies of those modes to excellent accuracy, with
relative differences of typically less than 5\%. For the simple MUn
EoS only in few cases the frequency of quasi-normal or nonlinear modes
could not be identified (marked by dashes in
Table~\ref{tab:frequencies_mun_eos}), presumably because of the
interaction with another sufficiently strongly excited mode nearby in
the spectrum, while for the microphysical EoS neither of the harmonics
can be extracted unambiguously (thus we refrain from presenting them
in Table~\ref{tab:frequencies_microphysical_eos}).

A comparison of the mode spectra of the migrating model US3 with those
of the corresponding perturbed post-migration equilibrium model PMS3
in Fig.~\ref{fig:spectrum_mun_eos} also exhibits that the collapse
associated with the dynamic migration in model US3 excites several
nonlinear couplings between quasi-normal modes, with some of them
being self-couplings like for instance the $ 2 \cdot F $-mode at twice
the frequency of the fundamental quasi-radial $ F $-mode.
\citet{abdikamalov_09_a} have also reported such strong nonlinear mode
excitation in their simulations of rotating neutron star models that
collapse to hybrid quark stars following a phase transition to quark
matter in the core. In contrast, in the only slightly perturbed model
PMS3 these nonlinear modes are very weak or even not present at all.
They make the same observation -- a strong presence of nonlinear modes
in the spectrum of the migration models and an efficient suppression
of such modes in the respective equilibrium models -- in all members
of the two model families, irrespective of the EoS used. We therefore
speculate that in a sufficiently strong dynamic gravitational
collapse, one can expect that nonlinear pulsations modes of
non-negligible amplitude are excited, independent of the physical
mechanism that activates the collapse.

\begin{table}
  \centering
  \caption{Frequencies for quasi-normal modes and some of their
    nonlinear self-couplings for the migration models of the family UM
    and the post-migration equilibrium models of the family PMM with
    the microphysical MUn EoS. $ f_F $ and $ f_{{}^{2\!}f} $ are the
    frequencies of the fundamental quasi-radial mode and the
    fundamental quadrupolar mode, while the other modes are nonlinear
    self-couplings. For the quasi-normal modes the relative
    differences in frequency between the model of the UM and PMM
    family are also given. For this model family, the strong damping
    of pulsations prohibits a clear identification of the harmonics.}
  \label{tab:frequencies_microphysical_eos}
  \begin{tabular}{@{}lc@{\quad}rc@{\quad}rcc@{}}
    \hline \\ [-1 em]
    Model &
    \multicolumn{2}{c}{$ f_F $} &
    \multicolumn{2}{c}{$ f_{{}^{2\!}f} $} &
    $ f_{2 \cdot\! F} $ &
    $ f_{3 \cdot\! F} $ \\
    &
    \multicolumn{2}{c}{[kHz]} &
    \multicolumn{2}{c}{[kHz]} &
    [kHz] &
    [kHz] \\ [0.2 em]
    \hline \\ [- 1 em]
    UM1  &
    1.061 &  \lowentry{9\%} &
    2.080 &  \lowentry{1\%} &
    2.122 &
    3.030
    \\
    PMM1 &
    1.156 &                &
    2.081 &                &
    2.286 &
    3.395
    \\ [0.5 em]
    UM2  &
    1.080 &  \lowentry{3\%} &
    2.100 &  \lowentry{1\%} &
    2.160 &
    3.161
    \\
    PMM2 &
    1.051 &                 &
    2.084 &                 &
    2.102 &
    3.153
    \\ [0.5 em]
    UM3  &
    1.000 & \lowentry{10\%} &
    2.139 &  \lowentry{2\%} &
    2.000 &
    3.031
    \\
    PMM3 &
    1.100 &                 &
    2.103 &                 &
    2.182 &
    3.280
    \\ [0.5 em]
    UM4  &
    0.960 &  \lowentry{1\%} &
    2.129 &  \lowentry{1\%} &
    1.920 &
    2.741
    \\
    PMM4 &
    0.966 &                 &
    2.112 &                 &
    1.941 &
    2.701
    \\ [0.5 em]
    UM5  &
    0.942 & \lowentry{13\%} &
    2.141 &  \lowentry{1\%} &
    1.884 &
    2.882
    \\
    PMM5 &
    1.061 &                 &
    2.122 &                 &
    2.143 &
    3.222
    \\ [0.5 em]
    UM6  &
    0.845 &  \lowentry{2\%} &
    2.120 &  \lowentry{2\%} &
    1.700 &
    2.521
    \\
    PMM6 &
    0.824 &                 &
    2.080 &                 &
    1.660 &
    2.449
    \\
    \hline
  \end{tabular}
\end{table}

\begin{figure*}
  \centerline{\includegraphics[width = 180 mm]{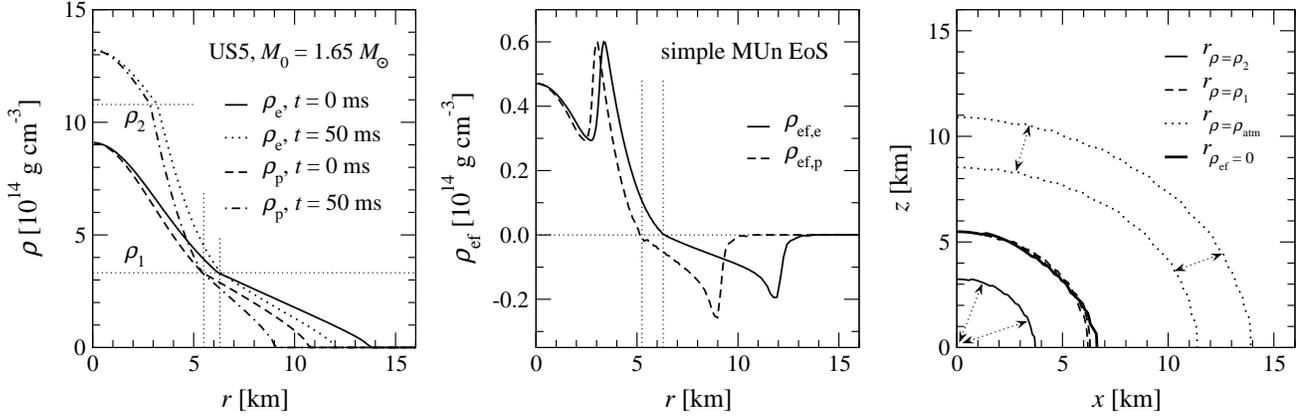}}
  \caption{Density topology of the most rapidly rotating migration
    model US5 of family US with the simple MUn EoS. In the left
    panel the radial profiles of the densities $ \rho_\mathrm{e} $ in
    the equatorial plane and $ \rho_\mathrm{p} $ along the polar axis
    are shown at the initial and final evolution time, respectively.
    The horizontal dotted lines mark the transition densities
    $ \rho_1 $ and $ \rho_2 $, while the vertical dotted lines specify
    the radius where $ \rho_\mathrm{e} $ and $ \rho_\mathrm{p} $ equal
    $ \rho_1 $ apparently independent of the evolution time. In the
    center panel the $ F $-mode density eigenfunction
    $ \rho_{F\mathrm{\!,e}} $ in the equatorial plane and
    $ \rho_{F\mathrm{\!,p}} $ along the polar axis is plotted. The vertical
    dotted lines indicate the respective nodes. In the right panel we
    show the range (symbolized by arrows) for the radial location of
    the surface where the density $ \rho $ equals $ \rho_2 $,
    $ \rho_1 $, and the atmosphere cutoff density
    $ \rho_\mathrm{atm} $ during the evolution, as well as the nodal
    surface of $ \rho_F $. Here $ x = r \, \sin \theta $ and
    $ z = r \, \cos \theta $ are the Cartesian coordinates.}
  \label{fig:eigenfunction_mun_eos}
\end{figure*}

A peculiar property of all models of family US with the simple MUn
EoS is revealed by a comparison of the evolution of their density
structure during the migration and the ring-down with the extracted
density eigenfunction $ \rho_F $ of the $ F $-mode, which is always
the mode that contains the largest pulsation energy (as e.g.\ shown by
the strong peak in the spectrum of $ \rho_\mathrm{c} $ in the left
panel of Fig.~\ref{fig:spectrum_mun_eos}). We observe that, as
expected, the density in the migrating neutron star changes during the
evolution everywhere in the star. Surprisingly, however, the density
remains practically constant at the two-dimensional surface where
$ \rho = \rho_1 $ initially, as demonstrated for the two radial
density profiles of model US5 along the equatorial plane and the polar
axis in the left panel of Fig.~\ref{fig:eigenfunction_mun_eos},
presented at both $ t = 0 $ and $ 50 \mathrm{\ ms} $. Apparently,
while $ \rho_\mathrm{c} $ jumps from well below to well above
$ \rho_2 $, the (angular dependent) location where $ \rho = \rho_1 $
stays at almost the same radius. The plot of the corresponding density
eigenfunction profiles in the center panel of
Fig.~\ref{fig:eigenfunction_mun_eos} then shows that $ \rho_F $ has a
node virtually at the same location. In the right panel of
Fig.~\ref{fig:eigenfunction_mun_eos} this behavior is summarized by
plotting the maximum range during the entire evolution time of
(rotationally symmetric) surfaces where $ \rho = \rho_2 $ (closest to
the center, with a large variation and a minimum range of $ r = 0 $ as
$ \rho < \rho_2 $ initially in the star), where $ \rho = \rho_1 $
(which lies in a very narrow interval), and where
$ \rho = \rho_\mathrm{atm} $ (i.e.\ the boundary of the star which
again changes its extent significantly). The close proximity of the
surface with $ \rho = \rho_1 $ to the nodal surface of the
eigenfunction $ \rho_F $ is striking. We speculate that if nodal
surface is ``pinned down'' where the EoS qualitatively changes its
properties at the density $ \rho_1 $, which is also the location where
$ \partial P / \partial \rho $ becomes discontinuous. Apart from this
attempt of explanation we are not able to give a compelling reason for
this property of the US models, but it is obvious that regarding the
main pulsation mode the migrating star decomposes into two separate
domains with an interface at $ \rho = \rho_1 $, each of them pulsating
coherently, but with opposite phase.

Similarly, we find that the models of family UM with a microphysical
EoS exhibit a nodal surface for the $ F $-mode that is approximately
constant in space throughout the migration and the subsequent
ring-down, although here the surface expands a bit during the first
phase of the migration. However, the (again approximately constant)
density in the neutron star model corresponding to this nodal surface
now is not a distinguished location in the microphysical EoS, being
for instance significantly smaller than $ \rho_1 $, where the domain
with zero pressure slope starts. Unlike in the models with the simple
MUn EoS, in the models with the microphysical EoS the eigenfunction
topology is apparently not influenced by any transition density of the
EoS.

In passing, we point out that in contrast to the rotating neutron star
models by \citet{lin_06_a} that undergo a mini-collapse due to a
phase transition in the EoS leading to a formation of a quark matter
core, our models do not develop significant convection after the
initial contraction. This would be easily observable both as a
strong low-frequency contribution much below $ 1 \mathrm{\ kHz} $ in
the spectrum shown in Fig.~\ref{fig:spectrum_mun_eos} and as a
step-like pattern corresponding to the convection cells in the radial
rotation velocity profiles in
Figs.~\ref{fig:rotation_velocity_profile_mun_eos}
and~\ref{fig:rotation_velocity_profile_microphysical_eos}.
\citet{abdikamalov_09_a} have shown that for such a
phase-transition-induced collapse with the particular EoS treatment of
\citet{lin_06_a} the convection is caused by a significant negative
entropy gradient, which is not present in our models.

%%%%%%%%%%%%%%%%%%%%%%%%%%%%%%%%%%%%%%%%%%%%%%%%%%%%%%%%%%%%%%%%%%%%%%%%%%%%%%%%%%%
%%%%%%%%%%%%%%%%%%%%%%%%%%%%%%%%%%%%%%%%%%%%%%%%%%%%%%%%%%%%%%%%%%%%%%%%%%%%%%%%%%%

\subsection{Emission of gravitational waves and mode resonance}
\label{subsection:gravitational_wave_emission}

\begin{figure*}
  \centerline{\includegraphics[width = 180 mm]{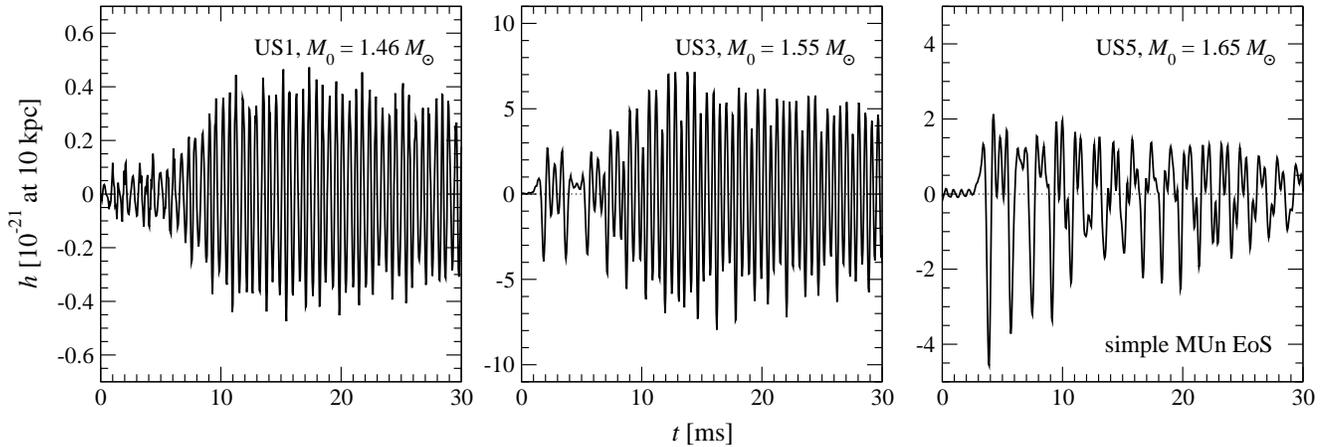}}
  \caption{Time evolution of the gravitational wave strain $ h $ for
    three migration models of family US with the simple MUn EoS.}
  \label{fig:waveform_mun_eos}
\end{figure*}

\begin{figure*}
  \centerline{\includegraphics[width = 180 mm]{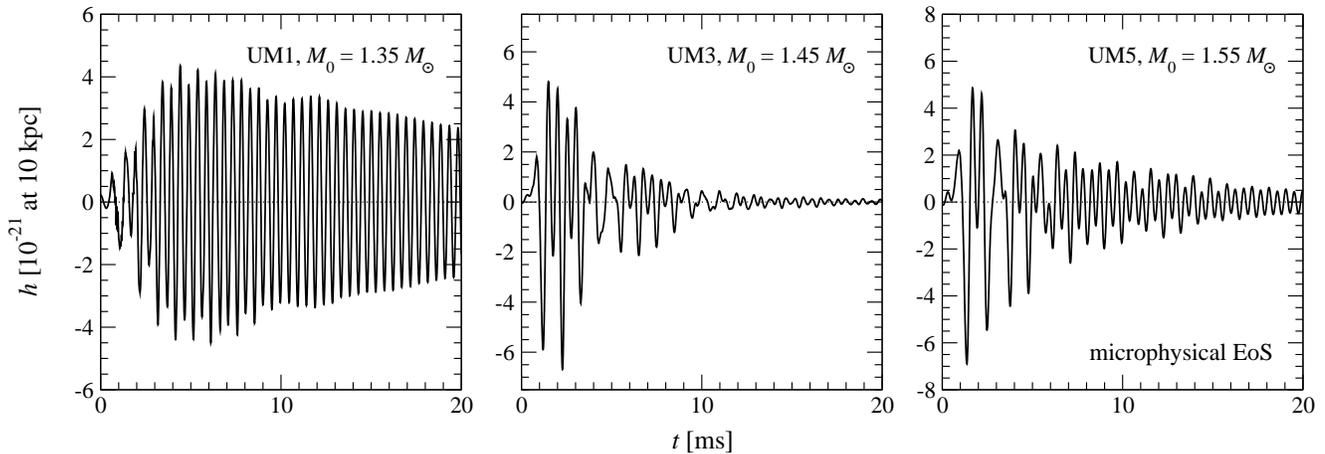}}
  \caption{Time evolution of the gravitational wave strain $ h $ for
    three migration models of family UM with the microphysical EoS.}
  \label{fig:waveform_microphysical_eos}
\end{figure*}

While the increase in density resulting from the migration is quite
uniform for all members of each model family, the strength of
gravitational wave emission, which is sensitive to the rotation
frequency, is expected to vary as the initial value of $ f $ and the
change $ \Delta f $ during the collapse varies among different models.
The dependence of the waveform on the rotation rate is clearly
revealed in the gravitational radiation waveform plots presented in
Figs.~\ref{fig:waveform_mun_eos}
and~\ref{fig:waveform_microphysical_eos} for three selected models of
each of the two families US and UM, respectively. The according
(absolute) peak amplitudes $ |h|_\mathrm{max} $ of the waveforms are
summarized in Table~\ref{tab:migration_dynamics}. The main emission
modes are the $ {}^{2\!}f $-mode and, for rapidly rotating models,
also the $ F $-mode.

\begin{figure}
  \centerline{\includegraphics[width = 85 mm]{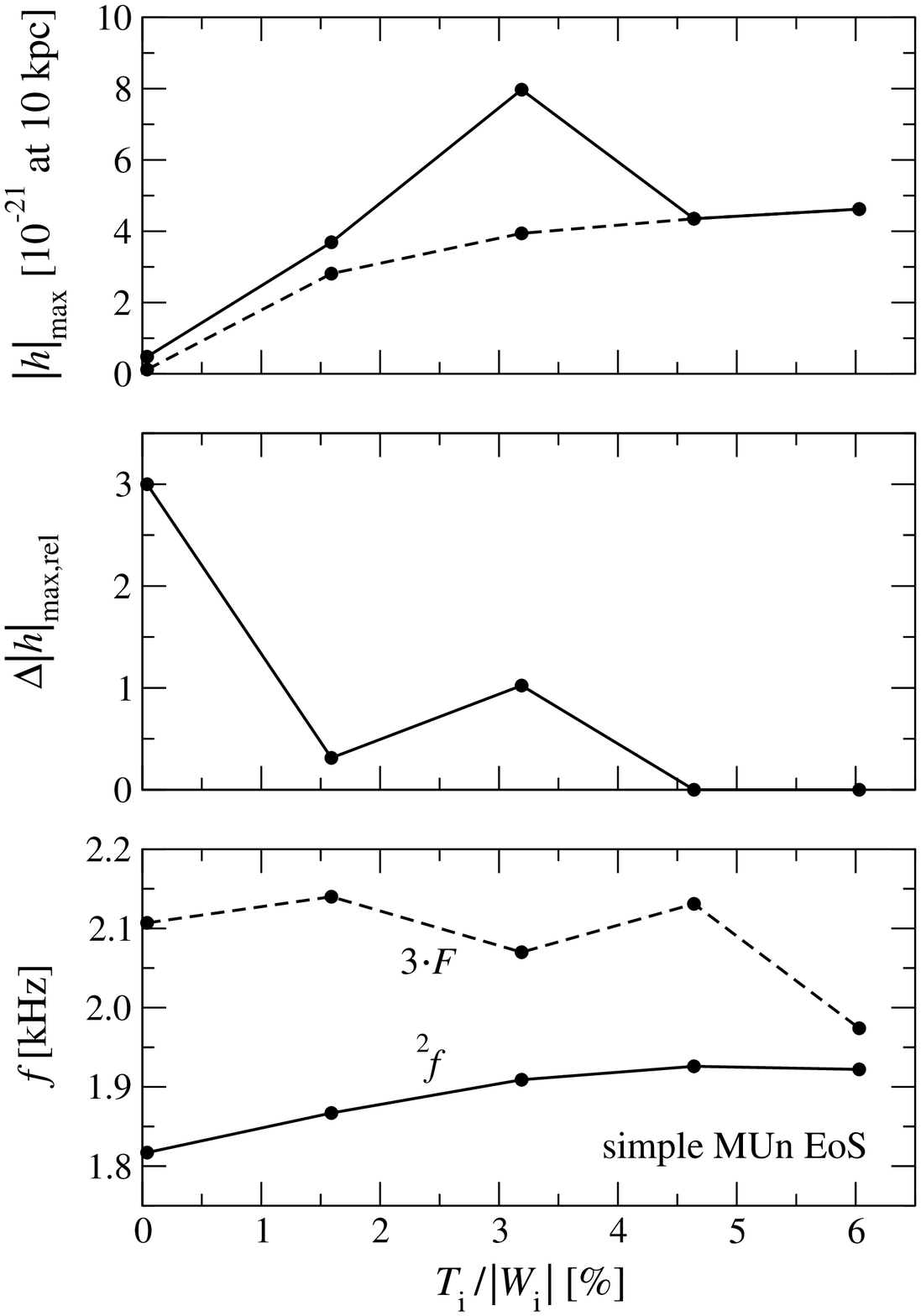}}
  \caption{Dependence of the maximum gravitational wave strain
    $ |h|_\mathrm{max} $ (top panel), the relative amplification of
    the maximum gravitational wave strain
    $ \Delta |h|_\mathrm{max,rel} $ (center panel), and the frequency
    $ f $ for potentially resonating modes (bottom panel) on the
    initial rotation rate $ T_\mathrm{i} / |W_\mathrm{i}| $ for the
    migration models of family US with the simple MUn EoS.
    $ |h|_\mathrm{max} $ is given with (solid line) and without
    (dashed line) taking into account the effects of mode resonance,
    and $ \Delta |h|_\mathrm{max,rel} $ is the relative increase in
    $ |h|_\mathrm{max} $ due to mode resonance. The frequencies of the
    $ {}^{2\!}f $-mode and $ 3 \cdot F $-mode are marked with a solid
    and dashed line, respectively.}
  \label{fig:resonance_mun_eos}
\end{figure}

\begin{figure}
  \centerline{\includegraphics[width = 85 mm]{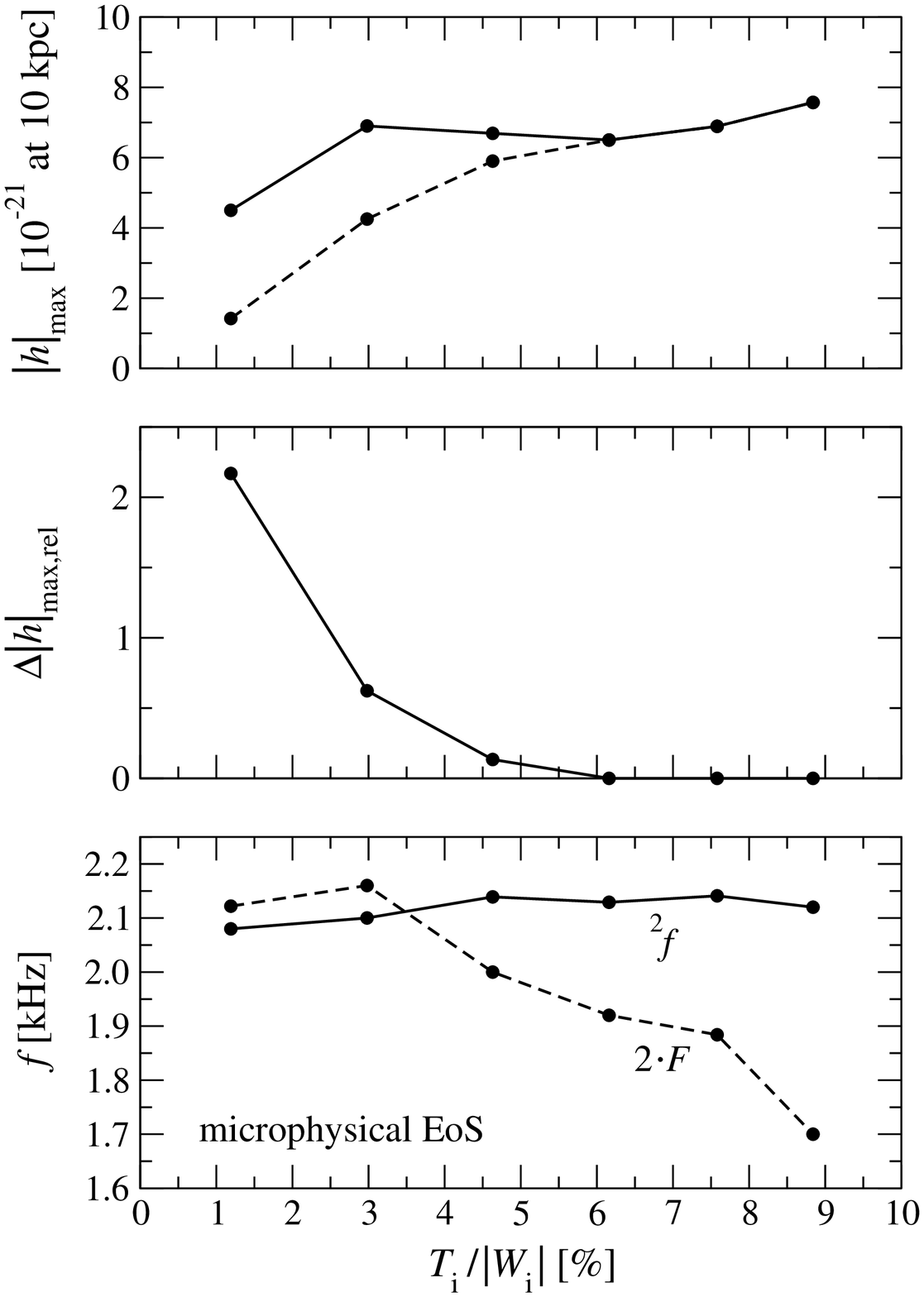}}
  \caption{Dependence of the maximum gravitational wave strain
    $ |h|_\mathrm{max} $ (top panel), the relative amplification of
    the maximum gravitational wave strain
    $ \Delta |h|_\mathrm{max,rel} $ (center panel), and the frequency
    $ f $ for potentially resonating modes (bottom panel) on the
    initial rotation rate $ T_\mathrm{i} / |W_\mathrm{i}| $ for the
    migration models of family UM with the microphysical EoS.
    $ |h|_\mathrm{max} $ is given with (solid line) and without
    (dashed line) taking into account the effects of mode resonance,
    and $ \Delta |h|_\mathrm{max,rel} $ is the relative increase in
    $ |h|_\mathrm{max} $ due to mode resonance. The frequencies of the
    $ {}^{2\!}f $-mode and $ 2 \cdot F $-mode are marked with a solid
    and dashed line, respectively.}
  \label{fig:resonance_microphysical_eos}
\end{figure}

It is remarkable for this migration scenario that $ |h|_\mathrm{max} $
apparently does not monotonically depend on either $ f_\mathrm{i} $ or
$ T_\mathrm{i} / |W_\mathrm{i}| $ (see e.g.\ the plot in the top
panels of Fig.~\ref{fig:resonance_mun_eos} for model family US and
Fig.~\ref{fig:resonance_microphysical_eos} for model family UM). An
instructive example is also the relatively slowly rotating model US1
(see left panel of Fig.~\ref{fig:waveform_mun_eos}), where the maximum
waveform amplitude during the initial collapse phase of the migration,
which lasts at most $ 2 \mathrm{\ ms} $, is around $ 10^{-22} $ at a
distance of $ 10 \mathrm{\ kpc} $ from the source. Considerably later,
at $ t \sim 5 \mathrm{\ ms} $, already well in the ring-down phase,
the amplitude gradually increases until it saturates at times after
$ \sim 10 \mathrm{\ ms} $ with a maximum
$ |h|_\mathrm{max} \sim 5 \times 10^{-22} $. Finally, after that the
amplitude of the quasi-periodic waveform's envelope decays as
expected (see also the qualitatively similar waveform of model UM1 in
the left panel of Fig.~\ref{fig:waveform_microphysical_eos}).

A very similar behavior was also found in recent simulations of
rotating neutron star models which collapse due to an EoS phase
transition \citep{abdikamalov_09_a}. In that work it was shown that
nonlinear modes can vigorously boost the emission of gravitational
waves if they are in resonance with an efficiently emitting
quadrupolar mode like the $ {}^{2\!}f $-mode, provided that both the
frequencies of those two modes are in close proximity and the
nonlinear mode contains a substantial amount of pulsation kinetic
energy \citep[see also][]{eardley_83_a, dimmelmeier_06_a}. While in
the models of \citet{abdikamalov_09_a} the nonlinear mode which drives
the mode resonance is the $ 2 \cdot F $-mode, the mode to be
considered here is the $ 3 \cdot F $-mode for the models with the
simple MUn EoS and the $ 2 \cdot F $-mode for the ones with the
microphysical EoS, as suggested by the frequencies given in
Tables~\ref{tab:frequencies_mun_eos}
and~\ref{tab:frequencies_microphysical_eos}, respectively.

Indeed Figs.~\ref{fig:resonance_mun_eos}
and~\ref{fig:resonance_microphysical_eos} show the connection between
the proximity of these potentially resonating modes and the maximum
waveform amplitude $ |h|_\mathrm{max} $ for all migrating models
considered here. By evaluating $ |h|_\mathrm{max} $ on the one hand
only in a time window close to the time of maximum compression (i.e.\
at the end of the dynamic initial collapse phase before mode resonance
can develop; dashed line) and on the other hand for the full waveform
(solid line), the enhancement of gravitational wave emission by
resonance can be quantified. For the models with a microphysical EoS
of family UM in Fig.~\ref{fig:resonance_microphysical_eos}, both the
absolute and the relative amplification of the waveform amplitude (see
top and center panel, respectively) is clearly strongest when the
$ 2 \cdot F $-mode is closest in frequency to the $ {}^{2\!}f $-mode.
This happens at low rotation, while for rapid rotation we detect no
imprint of mode resonance in the waveform and $ |h|_\mathrm{max} $
obtained close to the time of maximum compression is the global
maximum. The models with a simple MUn EoS presented in
Fig.~\ref{fig:resonance_mun_eos} reveal a slightly more complicated
behavior. Again the slowest rotator exhibits the strongest relative
increase of $ |h|_\mathrm{max} $ by mode resonance (see center panel),
albeit from a very small absolute level (see top panel). As rotation
grows and the frequencies of the $ 2 \cdot F $-mode and the
$ 3 \cdot F $-mode approach each other, the effects of the resonance
on $ |h|_\mathrm{max} $ become apparent. However, in the rapid
rotation limit mode resonance again ceases (as for the models of
family UM), although in family US the frequencies of the two relevant
modes are now in close proximity. Here the waveform again possess a
regularly damped sinusoidal pattern.

For such rapidly rotating models, we find that the peaks associated to
the $ {}^{2\!}f $-mode and the $ 3 \cdot F $-mode (for family US) or
the $ 2 \cdot F $-mode (for family UM) are clearly separated also
during the interval from $ 5 $ to $ 15 \mathrm{\ ms} $. Additionally,
the respective nonlinear self-coupling of the $ F $-mode is excited to
a much lesser degree than in slower rotating models with mode
resonance, and thus carries less transferable energy. Furthermore,
rapid rotation also causes the regular fundamental quasi-radial
$ F $-mode to acquire such a strong quadrupolar nature that it becomes
the main gravitational wave emitting mode, surpassing the
$ {}^{2\!}f $-mode in peak height in the waveform spectrum (see also
the discussion in
Section~\ref{subsection:gravitational_wave_detectability}). Therefore,
even if for such models the $ {}^{2\!}f $-mode was somewhat amplified,
this would not be identifiable in the spectrum, which is dominated by
the $ F $-mode. We also note that the inefficacy of this type of mode
resonance for rapid rotation was also found by
\citet{abdikamalov_09_a} in their simulations of the collapse of
neutron star models to hybrid quark stars.

We also note that although in the dynamic collapse phase of a
migration the self-couplings of the $ F $-mode, which are responsible
for transferring energy into the emitting $ {}^{2\!}f $-mode, are much
more violently excited, they are also present in perturbed equilibrium
models. Consequently, we observe the effects of mode resonance not
only in some of the migrating models of family US and UM, but also in
the corresponding equilibrium models of family PMS and PMM, if these
are subjected to a small amplitude initial perturbation according to
Eq.~(\ref{eq:trial_eigenfunction_0}), which favors the excitation of
quasi-radial $ l = 0 $ modes. As expected, the quadrupolar $ l = 2 $
perturbation of Eq.~(\ref{eq:trial_eigenfunction_2}) excites
self-couplings of the $ F $-mode only weakly, and thus in that case
mode resonance does not occur in any of the equilibrium models.

%%%%%%%%%%%%%%%%%%%%%%%%%%%%%%%%%%%%%%%%%%%%%%%%%%%%%%%%%%%%%%%%%%%%%%%%%%%%%%%%%%%
%%%%%%%%%%%%%%%%%%%%%%%%%%%%%%%%%%%%%%%%%%%%%%%%%%%%%%%%%%%%%%%%%%%%%%%%%%%%%%%%%%%

\subsection{Detectability prospects of gravitational waves}
\label{subsection:gravitational_wave_detectability}

Due to the long quasi-periodic emission time the gravitational waves
emitted by a migration event is a promising source for gravitational
wave detectors of both interferometer or resonant type. Assuming that
the strong post-collapse oscillations are not damped by any other
physical mechanism than dissipation of kinetic energy by shocks, the
damping we observe for the models of family US suggests that the
effective total emission time for gravitational waves can be much
longer than the evolution time of our models (which is
$ t_\mathrm{f} = 50 \mathrm{\ ms} $), i.e.\ last for hundreds of
pulsation cycles. In contrast, the models of family UM are so
strongly damped that the emission of gravitational waves has
essentially died out at the end of the evolution time at
$ t = 50 \mathrm{\ ms} $.

\begin{table*}
  \centering
  \caption{Detection prospects for gravitational waves.
    $ f_\mathrm{c} $ is the characteristic frequency, $ h_\mathrm{c} $
    is the integrated characteristic gravitational wave signal strain
    and $ \mathrm{SNR} $ is the signal-to-noise ratio, each given for
    the LIGO detector, the VIRGO detector, and the Advanced LIGO
    detector in broadband mode. All quantities assume a total emission
    time of $ t_\mathrm{f} = 50 \mathrm{\ ms} $ and are dependent on
    the rms strain noise $ h_\mathrm{rms} $ of the detector.}
  \label{tab:gravitational_wave_emission}
  \begin{tabular}{@{}lc@{~~}c@{~~}cc@{~~}c@{~~}cc@{~~}c@{~~}c@{}}
    \hline \\ [-1 em]
    Model &
    $ f_\mathrm{c,LIGO} $ &
    $ f_\mathrm{c,VIRGO} $ &
    $ f_\mathrm{c,Adv.\,LIGO} $ &
    $ h_\mathrm{c,LIGO} $ &
    $ h_\mathrm{c,VIRGO} $ &
    $ h_\mathrm{c,Adv.\,LIGO} $ &
    $ \mathrm{SNR}_\mathrm{LIGO} $ &
    $ \mathrm{SNR}_\mathrm{VIRGO} $ &
    $ \mathrm{SNR}_\mathrm{Adv.\,LIGO} $ \\
    &
    \highentry{[khz]} &
    \highentry{[khz]} &
    \highentry{[khz]} &
    $ \displaystyle \left[ \!\!\!
      \begin{array}{c}
        10^{-21} \\ [-0.2 em]
        \mathrm{\ at\ 10\ kpc}
      \end{array}
    \! \right] $ &
    $ \displaystyle \left[ \!\!\!
      \begin{array}{c}
        10^{-21} \\ [-0.2 em]
        \mathrm{\ at\ 10\ kpc}
      \end{array}
    \! \right] $ &
    $ \displaystyle \left[ \!\!\!
      \begin{array}{c}
        10^{-21} \\ [-0.2 em]
        \mathrm{\ at\ 10\ kpc}
      \end{array}
    \! \right] $ &
    \highentry{[at 10 kpc]} &
    \highentry{[at 10 kpc]} &
    \highentry{[at 10 kpc]} \\ [0.7 em]
    \hline \\ [-1 em]
    US1 & 1.768 & 1.786 & 1.798 & \z2.6 & \z2.6 & \z2.6 & 0.3 & 0.5 & \zz9.2 \\
    US2 & 1.225 & 1.321 & 1.432 &  10.4 &  11.0 &  11.7 & 2.1 & 3.4 & \z54.9 \\
    US3 & 1.574 & 1.651 & 1.728 &  31.6 &  32.9 &  34.1 & 4.4 & 7.6 &  127.6 \\
    US4 & 0.924 & 1.006 & 1.129 & \z9.3 & \z9.8 &  10.7 & 2.9 & 4.4 & \z65.0 \\
    US5 & 0.718 & 0.766 & 0.834 & \z5.8 & \z6.0 & \z6.4 & 2.6 & 3.7 & \z51.8 \\ [0.5 em]
    UM1 & 2.024 & 2.031 & 2.041 &  18.6 &  18.7 &  18.7 & 1.8 & 3.2 & \z56.6 \\
    UM2 & 1.928 & 1.952 & 1.983 &  22.0 &  22.2 &  22.5 & 2.3 & 4.0 & \z70.6 \\
    UM3 & 1.236 & 1.297 & 1.384 & \z7.0 & \z7.2 & \z7.6 & 1.4 & 2.3 & \z37.0 \\
    UM4 & 1.130 & 1.190 & 1.282 & \z7.9 & \z8.2 & \z8.6 & 1.8 & 2.9 & \z45.7 \\
    UM5 & 1.047 & 1.110 & 1.202 & \z7.5 & \z7.8 & \z8.3 & 1.9 & 3.1 & \z47.3 \\
    UM6 & 1.007 & 1.078 & 1.178 & \z8.8 & \z9.2 & \z9.8 & 2.4 & 3.8 & \z57.3 \\
    \hline
  \end{tabular}
\end{table*}

\begin{figure*}
  \centerline{\includegraphics[width = 180 mm]{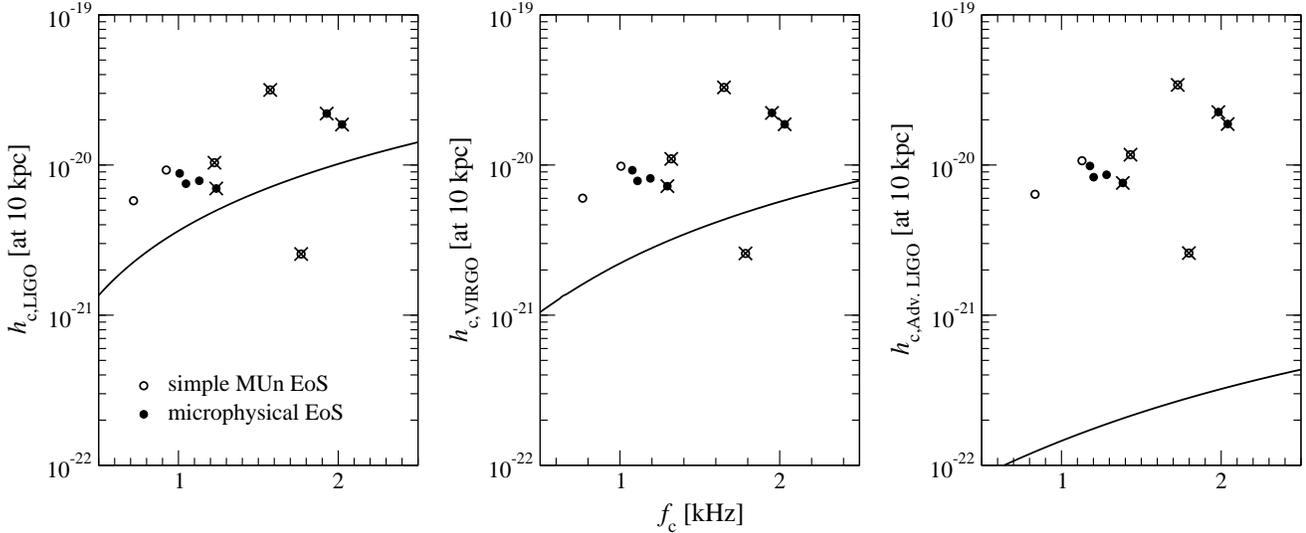}}
  \caption{Location of the gravitational wave signals for the
    migration models from family US with the simple MUn EoS (filled
    circles) and family UM with the microphysical EoS (open circles)
    in the $ h_\mathrm{c} $--$ f_\mathrm{c} $ plane relative to the
    sensitivity curves (i.e.\ the rms strain noise $ h_\mathrm{rms} $)
    of the LIGO detector (left panel), VIRGO detector (center panel),
    and Advanced LIGO detector in broadband mode (right panel), all at
    a distance of $ 10 \mathrm{\ kpc} $. Models whose gravitational
    wave emission is amplified by mode resonance effects are marked
    with a cross.}
  \label{fig:detector_sensitivity}
\end{figure*}

To assess the prospects for detection by current and planned
interferometric detectors, we calculate characteristic quantities for
the gravitational wave signal following \citet{thorne_87_a}. Applying
a Fourier transform to the dimensionless gravitational wave strain
$ h $, we obtain
\begin{equation}
  \hat{h} =
  \int_{-\infty}^\infty \!\! e^{2 \pi i f t} h \, dt.
  \label{eq:waveform_fourier_tranform}
\end{equation}
One can compute the (detector dependent) integrated characteristic
frequency as
\begin{equation}
  f_\mathrm{c} =
  \left( \int_0^\infty \!
  \frac{\langle \hat{h}^2 \rangle}{S_h}
  f \, df \right)
  \left( \int_0^\infty \!
  \frac{\langle \hat{h}^2 \rangle}{S_h}
  df \right)^{-1}\!\!\!\!\!\!\!,
  \label{eq:characteristic_frequency}
\end{equation}
and the dimensionless integrated characteristic strain as
\begin{equation}
  h_\mathrm{c} =
  \left( 3 \int_0^\infty \!
  \frac{S_{h\mathrm{\,c}}}{S_h} \langle \hat{h}^2 \rangle
  f \, df \right)^{1/2}\!\!\!\!\!\!\!\!,
  \label{eq:characteristic_amplitude}
\end{equation}
where $ S_h $ is the power spectral density of the detector and
$ S_{h\mathrm{\,c}} = S_h (f_\mathrm{c}) $. We approximate the average
$ \langle \hat{h}^2 \rangle $ over randomly distributed angles by
$ \hat{h}^2 $, assuming optimal orientation of the detector. From
Eqs.~(\ref{eq:characteristic_frequency},
\ref{eq:characteristic_amplitude}) the signal-to-noise ratio can be
calculated as
$ \mathrm{SNR} = h_\mathrm{c} / [h_\mathrm{rms} (f_\mathrm{c})] $,
where $ h_\mathrm{rms} = \sqrt{f S_h} $ is the value of the rms strain
noise for the detector (which gives the theoretical sensitivity
window). In Table~\ref{tab:gravitational_wave_emission} we summarize
the values of $ f_\mathrm{c} $, $ h_\mathrm{c} $ and the
signal-to-noise ratio for all the migrating models of the two families
US and UM for the currently operating LIGO and VIRGO detectors and for
the future Advanced LIGO detector in broad band operation
mode\footnote{Note that both in
  Table~\ref{tab:gravitational_wave_emission} and in
  Fig.~\ref{fig:detector_sensitivity} the gravitational wave
  characteristics have been evaluated for a total emission time of
  $ t_\mathrm{f} = 50 \mathrm{\ ms} $, where the waveform amplitude of
  at least the models of family US has not yet decayed to zero.
  Therefore, for these models the values for $ h_\mathrm{c} $ and the
  signal-to-noise ratio shown here are actually a lower bound.}. Here
we consider a source inside our own Galaxy at a reference distance of
$ 10 \mathrm{\ kpc} $.

The fraction of neutron stars that undergo a migration process at some
stage in their lifetimes is rather uncertain. In this work we
restrict ourselves to considering isolated neutron stars. Depending on
the selected gravitational mass and the EoS, such a neutron star must
have spun down to somewhere between $ 50 $ and $ 800 \mathrm{\ Hz} $
to be subject to a migration (see
Table~\ref{tab:unstable_initial_models}, and
Figs.~\ref{fig:back_bending_mun_eos}
and~\ref{fig:back_bending_microphysical_eos}), provided
it has an EoS that is similar to one of our choice and shows the back
bending phenomenon.

Apparently, isolated millisecond pulsars with low magnetic field,
which have been recycled by accretion, are no appropriate candidates,
as they spin down too slowly, on a Hubble time scale. However, there
exists a another class of neutron stars, which are thought to rotate
initially at a millisecond period, and those are magnetars. In order
to generate the required surface magnetic field of roughly $ 10^{14} $
to $ 10^{15} \mathrm{\ G} $, the proto-neutron star born in a type~II
supernova explosion must rotate at $ \sim 500 \mathrm{\ Hz} $. Due to
strong magnetic breaking, it slows down to a few Hz after only about
$ 10^3 \mathrm{\ y} $. 

The most dramatic spin-down of a magnetar occurs during the first
years of its life, which is the period when the migration phenomenon
could occur. So far ten magnetars were detected in Galaxy: six
anomalous X-ray pulsars and four soft-gamma repeaters. Magnetars are
active and therefore observable for about
$ 10^3\mbox{\,--\,}10^4 \mathrm{\ y} $. This means that their birth
rate is not much smaller than that of ``ordinary'' radio pulsars (with
a lifetime of $ \sim 10^6 \mathrm{\ y} $), which in turn is comparable
to the type II supernova rate in a big spiral galaxy like ours.
Consequently, we may expect about one migration event in our Galaxy
per century, which is a very low rate. Consistent with this simple
argument, an analysis based on observational data yields a very
uncertain birth rate between 0.2 and 6 per century \citep{muno_07_a,
  gill_07_a}. The magnetar birth rate rises to about one per year if
the whole Virgo cluster of galaxies is considered
\citep{dallosso_07_a}.

Still, if such a migration event did occur in our Galaxy, for current
interferometric detectors of the LIGO class and assuming an emission
time $ t_\mathrm{f} = 50 \mathrm{\ ms} $, all of our models except the
slowly rotating one US1 have a signal-to-noise ratio very close to or
above $ 1 $. Model US3, which is both moderately rapidly rotating and
whose gravitational wave emission is amplified by mode resonance (see
center panel of Fig.~\ref{fig:waveform_mun_eos}), features the largest
signal-to-noise ratio of all investigated models at about $ 4 $ for
current LIGO and $ 8 $ for current VIRGO. For the Advanced LIGO
detector, the signal-to-noise ratio lies comfortably above $ 10 $ for
all models (again except model US1 which still yields
$ \mathrm{SNR} = 9 $) and reaches $ \sim 130 $ in model US3. It is
remarkable that $ h_\mathrm{c} $ (and correspondingly, also the
signal-to-noise ratio) for the model family UM with the microphysical
EoS is bounded by a rather narrow interval between about $ 7 $ and
$ 22 \times 10^{-21} $ at $ 10 \mathrm{\ kpc} $, although the initial
rotation rates of the models span a wide range (see
Table~\ref{tab:unstable_initial_models}).

For substantially increasing the event rate, it would be necessary for
the detector to be sensitive to signals coming from distances out to
the Virgo cluster, at $ \sim 20 \mathrm{\ Mpc} $. However, at this
distance the signal-to-noise ratios for our models given in
Table~\ref{tab:gravitational_wave_emission} must to be scaled down by
a factor of $ 2000 $ and consequently drop to well below $ 1 $ even
for Advanced LIGO. Therefore, as for gravitational wave signals from
supernova core collapse \citep[see the discussion
in][]{dimmelmeier_07_a}, the second generation detectors will improve
the signal-to-noise ratio of a local event alone, but will not
increase the event rate much on account of the inhomogeneous galaxy
distribution in the local region of the Universe. Only third
generation detectors will have the required sensitivity in the kHz
range to achieve a robust signal-to-noise ratio at the distance of the
Virgo cluster.

Note that for most of the models the \emph{integrated} characteristic
frequency $ f_\mathrm{c} $ given in
Table~\ref{tab:gravitational_wave_emission} is not very close to
either of the two main gravitational wave emission frequencies $ f_F $
and $ f_{{}^{2\!}f} $. This is because $ f_\mathrm{c} $ reflects the
frequency dependence of the sensitivity of the detector, because
nonlinear mode couplings and higher-order linear modes also contribute
to the gravitational wave signal (although with lower relative
amplitudes) and, most importantly, because for many models the
gravitational wave power spectrum of the signal exhibits two strong
peaks from the $ F $-mode and the $ {}^{2\!}f $-mode. Based on the
peak height in the spectrum of the gravitational radiation waveform,
we find that the fundamental quadrupolar $ {}^{2\!}f $-mode is
the strongest emitter for the slowly and moderately rapidly rotating
models, followed by the fundamental quasi-radial $ F $-mode. For rapid
rotation, this order of importance is reversed.

The detector dependence of $ f_\mathrm{c} $ and $ h_\mathrm{c} $ is
also illustrated in Fig.~\ref{fig:detector_sensitivity}, where the
locations of the gravitational wave signals for all models are plotted
relative to the rms strain noise $ h_\mathrm{rms} $ of the current
LIGO and VIRGO detectors (left and center panels, respectively) and
the future Advanced LIGO detector (right panel), all for a distance to
the source of $ 10 \mathrm{\ kpc} $. Model US3, which both has
significant rotation and exhibits mode resonance (models where more
resonance amplifies gravitational radiation emission are marked by
crosses in Fig.~\ref{fig:detector_sensitivity}), clearly sticks out
with the largest values of $ h_\mathrm{c} $. In general, models with
initially moderate rotation, whose waveform amplitudes are in addition
boosted by mode resonance effects, are found in the upper right
quarter of the diagram, i.e.\ they feature both a high characteristic
frequency and a large characteristic strain. Model US1, which is by
far the slowest rotator, shows the smallest value for
$ h_\mathrm{c} $, albeit at a comparably high $ f_\mathrm{c} $.

From Fig.~\ref{fig:detector_sensitivity} it is clear that a possible
detection is hampered by the adverse characteristics of the detector
sensitivity curve in the frequency interval relevant for our models,
where the strain noise $ h_\mathrm{rms} $ rises steeply with
frequency. Proposed future detectors with a flat $ h_\mathrm{rms} $
curve up to frequencies in the kHz range could eliminate this problem.

\begin{figure}
  \centerline{\includegraphics[width = 85 mm]{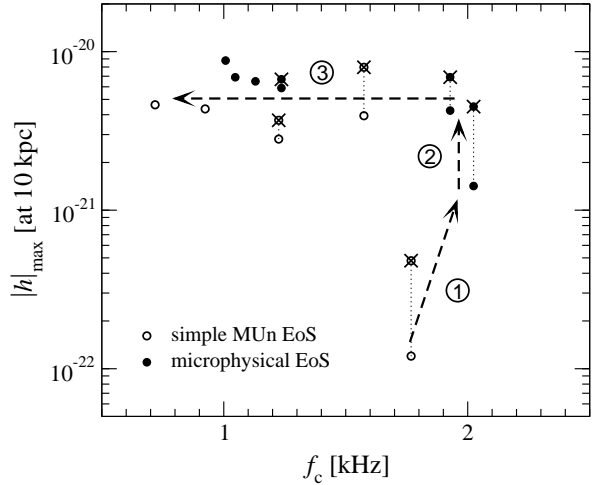}}
  \caption{Dependence of the maximum gravitational wave strain
    $ |h|_\mathrm{max} $ on the characteristic frequency
    $ f_\mathrm{c} $ for the migration models of family US with the
    simple MUn EoS (unfilled circles) and family UM with the
    microphysical EoS (filled circles). If mode resonance influences
    gravitational wave emission (models denoted with a cross),
    $ |h|_\mathrm{max} $ also is computed around the time of maximum
    compression before the amplification of $ h $ takes effect. The
    decrease of $ |h|_\mathrm{max} $ without resonance effects is
    marked by dotted lines. The arrows indicate the qualitative effect
    of increasing initial rotation rate
    $ T_\mathrm{i} / |W_\mathrm{i}| $.}
  \label{fig:detector_sensitivity_resonance}
\end{figure}

Interestingly, if a larger selection of models that are evenly
distributed in the parameter space of initial rotation rate were
simulated, their distribution in the
$ h_\mathrm{c} $--$ f_\mathrm{c} $ plane would most probably
very well approximate the pattern found by \citet{dimmelmeier_08_a}
for rotating models of collapsing stellar cores in a supernova event.
In order to filter out the effects of mode resonance, which for some
models amplifies the waveform amplitude in the ring-down following the
collapse phase and which cannot be simply subtracted from the
time integrated characteristic strain $ h_\mathrm{c} $, in
Fig.~\ref{fig:detector_sensitivity_resonance} instead of
$ h_\mathrm{c} $ we plot the maximum gravitational wave strain
$ |h|_\mathrm{max} $ around the time of maximum compression (i.e.\ the
most dynamic moment of the collapse phase) before any possible
waveform amplification by mode resonance.

As for $ h_\mathrm{c} $ in Fig.~17 of \citet{dimmelmeier_08_a},
starting from very slowly rotating models $ |h|_\mathrm{max} $ and
(not as strongly) $ f_\mathrm{c} $ increase as rotation grows (see
arrow 1 in their figure and our
Fig.~\ref{fig:detector_sensitivity_resonance}), then only
the waveform amplitude rises at practically constant $ f_\mathrm{c} $
(arrow 2 again in both figures). We recall that for these models the
$ {}^{2\!}f $-mode is the strongest emitter, which depends only weakly
on rotation \citep[see Tables~\ref{tab:frequencies_mun_eos}
and~\ref{tab:frequencies_microphysical_eos}, and
also][]{dimmelmeier_06_a}, which explains the approximate constancy of
$ f_\mathrm{c} $ in this regime. This weak imprint of rotation is in
stark contrast to e.g.\ a rigidly rotating bar, where the emission
frequency increased linearly with rotation frequency.

Finally, for rapidly rotating models which run into the centrifugal
barrier, $ |h|_\mathrm{max} $ saturates at a constant or even slightly
declining level with significantly decreasing frequency
$ f_\mathrm{c} $ (arrow 3 again in both figures). What we do not
observe here is the rather separate group of extremely rotating
models, whose evolution and waveforms are dominated by centrifugal
forces (marked by 4 in their figure). In addition, in our case mode
resonance effects would bulge out the ``knee'' for moderately rotating
models in the upper right corner of the curve towards higher values of
$ h_\mathrm{c} $ (or in our case $ |h|_\mathrm{max} $), as seen in
Fig.~\ref{fig:detector_sensitivity_resonance}, where for such models
the maximum amplitude for the \emph{entire} waveform (marked by
crosses) is higher than if only taken in the period around maximum
compression.

%%%%%%%%%%%%%%%%%%%%%%%%%%%%%%%%%%%%%%%%%%%%%%%%%%%%%%%%%%%%%%%%%%%%%%%%%%%%%%%%%%%
%%%%%%%%%%%%%%%%%%%%%%%%%%%%%%%%%%%%%%%%%%%%%%%%%%%%%%%%%%%%%%%%%%%%%%%%%%%%%%%%%%%
%%%%%%%%%%%%%%%%%%%%%%%%%%%%%%%%%%%%%%%%%%%%%%%%%%%%%%%%%%%%%%%%%%%%%%%%%%%%%%%%%%%

\section{Summary and conclusions}
\label{section:conclusions}

We have studied the mini-collapse of rotating neutron star models
induced by a dynamic back bending instability. This instability is
caused by the softening of the equation of state resulting from a
phase transition in the dense core. The collapse leads to a migration
of the neutron star from an unstable equilibrium configuration to a
stable one after a series of damped pulsations. The migration starts
when the configuration reaches the local minimum of the total angular
momentum in the back bending diagram. We have investigated two
families of models, one with the simple analytic MUn equation of state
and one with a tabulated microphysical equation of state that includes
effects of a phase transition due to kaon condensation.

If the rotation profile of the neutron star remained uniform during
the collapse, the neutron star would simply migrate horizontally
(i.e.\ at constant total angular momentum) from the unstable branch to
the stable branch (which are separated by the unstable segment) in the
angular momentum vs.\ rotation frequency diagram. However, our
simulations show that the neutron star actually develops a
differential rotation profile. Still, we find that the final degree of
differential rotation is not large and that the new equilibrium
configuration is similar to the corresponding one on the stable branch
in the back bending diagram under the assumption of rigid rotation.

The dynamic collapse in the migration process creates a whole spectrum
of coupled pulsation modes, which we extract directly from the
numerical results of our fully nonlinear simulations. In addition, we
have also constructed equilibrium models which closely resemble the
post-migration configurations, and evolved them after applying a small
perturbation. Using both approaches, we have identified the
characteristics of the main modes excited by the mini-collapse, and
find evidence for resonance between the fundamental quadrupolar mode
and nonlinear self-couplings of the fundamental quasi-radial mode. Our
results show that the amplitude and waveform of gravitational
radiation depends strongly on the mass of the star and the equation of
state used. Namely, models with the microphysical of state, which
features a density jump associated with a first-order phase
transition, suffer from much stronger damping than models with the
mixed-phase MUn equation of state.

In the case of a detection with current or future gravitational wave
interferometer detectors (like VIRGO or LIGO), such imprints of the
model parameters on the waveform properties could be exploited to
observationally constrain the unknown equation of state for dense
matter. For a Galactic event, whose frequency of occurrence should be
comparable with the local type~II supernova rate, the predicted
gravitational wave signal from such a dynamic migration lies above the
designed sensitivity curves of current km-size detectors. With future,
more sensitive detectors, signals from events within the Virgo cluster
of galaxies could also come into range. Then the corresponding event
rate could possibly increase to about one per year. In addition, we
find that in some cases the gravitational wave emission is
considerably enhanced due to the said mode resonance, which transfers
energy from the strongly excited but weakly radiating quasi-radial
mode to the initially weakly excited but strongly radiating
quadrupolar mode.

Particularly promising candidates for a dynamic migration are young
magnetars, as they have an initial rotation rate in the kHz range
and spin down very rapidly. While the prospective astrophysical
phenomenon we study in this work seems to be rare, it has nevertheless
a potential of producing a very characteristic signature in the form
of a peculiar burst of gravitational radiation. The damping of the
post-migration pulsations, which is directly reflected in the
gravitational wave signal, is even sensitive to the character of the
phase transition softening, i.e.\ whether it occurs with or without a
jump in density. Therefore, if chance allows us to detect such a
dynamic migration event in our Galaxy, this will possibly be a source
of rich information about the structure of the superdense neutron star
core.

We point out that in our simulations the effects of the magnetic
field, which for magnetars has a typical strength of
$ B \sim 10^{14}\mbox{\,--\,}10^{15} \mathrm{\ G} $, are neglected.
In particular the impact on the stellar shape is discarded. However,
the typical pressure in neutron star cores, where the phase transition
in the equation of state that causes the back bending phenomenon
occurs, is $ 10^{34}\mbox{\,--\,}10^{35} \mathrm{\ dyn\ cm}^{-2} $
(see Figure~\ref{fig:equation_of_state}), while with $ B^2 / (8 \pi)
\sim 4 \times 10^{29} (B / 10^{15} \mathrm{\ G})^2 $ the stress
due to the magnetic field is many orders of magnitude lower.
Therefore, the influence of even a strong magnetic field on the core
structure is small, and consequently we expect only secondary effects
on the back bending instability.

We close by pointing out that the results presented here should be
considered as a first step towards a better understanding of the
neutron star interior and dynamical phenomena like migration from
unstable to stable configurations. Due to the presently very limited
knowledge about the equation of state for dense matter, some of the
aspects of the physical processes underlying the investigated scenario
are necessarily simplified. The treatment of the equation of state as
a representation of matter in the mixed phase experiencing a phase
transition is rather crude, especially when one considers the behavior
of fluid elements undergoing successive stages of compression and
decompression, and changing the proportions of the phases. In
particular, our modeling neglects any possible effects from local
heating, or from the creation and subsequent emission of neutrinos
during the migration collapse and the subsequent bounces. On the other
hand we have shown that the time scale of the mini-collapse is short
enough to neglect non-equilibrium and bulk viscosity effects.

%%%%%%%%%%%%%%%%%%%%%%%%%%%%%%%%%%%%%%%%%%%%%%%%%%%%%%%%%%%%%%%%%%%%%%%%%%%%%%%%%%%
%%%%%%%%%%%%%%%%%%%%%%%%%%%%%%%%%%%%%%%%%%%%%%%%%%%%%%%%%%%%%%%%%%%%%%%%%%%%%%%%%%%
%%%%%%%%%%%%%%%%%%%%%%%%%%%%%%%%%%%%%%%%%%%%%%%%%%%%%%%%%%%%%%%%%%%%%%%%%%%%%%%%%%%

\section*{Acknowledgments}

It is a pleasure to thank J\'er\^ome Novak, Christian D.\ Ott and
Nikolaos Stergioulas for helpful comments and discussions. Part of
this work was done during a visit of H.D.\ at the Nicolaus Copernicus
Astronomical Center in Warsaw, Poland. He expresses his gratitude for
the hospitality of the host group. H.D.\ acknowledges a Marie Curie
Intra-European Fellowship within the 6th European Community Framework
Programme (IEF 040464), and M.B.\ acknowledges Marie Curie Fellowships
within the 6th and 7th European Community Framework Programmes
(MEIF-CT-2005-023644 and ERG-2007-224793). This work was supported by
the DAAD and IKY (IKYDA German--Greek research travel grant) and by
the Polish MNiSW grant no.\ N20300632/0450.

%%%%%%%%%%%%%%%%%%%%%%%%%%%%%%%%%%%%%%%%%%%%%%%%%%%%%%%%%%%%%%%%%%%%%%%%%%%%%%%%%%%
%%%%%%%%%%%%%%%%%%%%%%%%%%%%%%%%%%%%%%%%%%%%%%%%%%%%%%%%%%%%%%%%%%%%%%%%%%%%%%%%%%%
%%%%%%%%%%%%%%%%%%%%%%%%%%%%%%%%%%%%%%%%%%%%%%%%%%%%%%%%%%%%%%%%%%%%%%%%%%%%%%%%%%%

%%%%%%%%%%%%%%%%%%%%%%%%%%%%%%%%%%%%%%%%%%%%%%%%%%%%%%%%%%%%%%%%%%%%%%%%%%%%%%%%%%%
%%%%%%%%%%%%%%%%%%%%%%%%%%%%%%%%%%%%%%%%%%%%%%%%%%%%%%%%%%%%%%%%%%%%%%%%%%%%%%%%%%%
%%%%%%%%%%%%%%%%%%%%%%%%%%%%%%%%%%%%%%%%%%%%%%%%%%%%%%%%%%%%%%%%%%%%%%%%%%%%%%%%%%%

\appendix

%%%%%%%%%%%%%%%%%%%%%%%%%%%%%%%%%%%%%%%%%%%%%%%%%%%%%%%%%%%%%%%%%%%%%%%%%%%%%%%%%%%
%%%%%%%%%%%%%%%%%%%%%%%%%%%%%%%%%%%%%%%%%%%%%%%%%%%%%%%%%%%%%%%%%%%%%%%%%%%%%%%%%%%
%%%%%%%%%%%%%%%%%%%%%%%%%%%%%%%%%%%%%%%%%%%%%%%%%%%%%%%%%%%%%%%%%%%%%%%%%%%%%%%%%%%

\section{Non-equilibrium effects in the equations of state}
\label{appendix:nonequilibrium_effects}

In its quasi-stationary evolution phase, the rotating neutron star
slowly proceeds along the stable branch. Here the slow angular
momentum loss results in an according compression of the neutron star
matter. When it enters the dynamic migration period, the phase
transition in the stellar core leads to a much more rapid initial
contraction, followed by several re-expansion and compression cycles
(see Figs.~\ref{fig:density_evolution_mun_eos}
and~\ref{fig:density_evolution_microphysical_eos}). In both regimes
the composition of matter has to re-adjust to the change in density
and pressure to keep matter in thermodynamic equilibrium. This process
of equilibration of composition proceeds via nuclear reactions, of
which the ones involving weak interaction (changing lepton number
and/or strangeness per baryon) are the slowest. We now investigate
these relaxation processes and their impact on the EoS of compressed
matter separately for the two regimes.

When the neutron star evolves on the stable branch (marked by solid
lines in Figs.~\ref{fig:back_bending_mun_eos}
and~\ref{fig:back_bending_microphysical_eos}), it move slowly
downwards along the branch on a characteristic secular time scale
$ \tau_\mathrm{sec} = \rho_\mathrm{c} / \dot{\rho}_\mathrm{c} \propto
(J / |\dot{J}|)^{-1} $. This time scale is so long that equilibration
nearly catches up with compression, and any deviations from
equilibrium are tiny. Consequently, the EoS of matter can be well
approximated by assuming full equilibrium.

As soon as the marginally stably equilibrium configuration is reached,
further loss of $ J $ triggers the dynamic migration, associated with
a compression of matter on the dynamic time scale
$ \tau_\mathrm{dyn} \lesssim 1 \mathrm{\ ms} $. Now matter moves off
weak equilibrium and the actual EoS stiffens compared to the fully
equilibrated one \citep[see e.g.][and references
therein]{gourgoulhon_95_a, haensel_02_a}. However, a deviation from
equilibrium accelerates (in a strongly nonlinear manner) the rate of
equilibration processes, which tend to return the matter towards
equilibrium following le Ch\^atelier's principle. Consequently, also
in the migration phase (i.e.\ the rapid initial contraction and the
subsequent pulsations) the EoS is quite close to the equilibrium one.
Strictly speaking, the deviation from equilibrium is such that as to
make the rate of equilibration equal to the time scale of dynamic
compression. Still, the effect on the dynamics is small, and is
thus being neglected in our calculation.

%%%%%%%%%%%%%%%%%%%%%%%%%%%%%%%%%%%%%%%%%%%%%%%%%%%%%%%%%%%%%%%%%%%%%%%%%%%%%%%%%%%
%%%%%%%%%%%%%%%%%%%%%%%%%%%%%%%%%%%%%%%%%%%%%%%%%%%%%%%%%%%%%%%%%%%%%%%%%%%%%%%%%%%
%%%%%%%%%%%%%%%%%%%%%%%%%%%%%%%%%%%%%%%%%%%%%%%%%%%%%%%%%%%%%%%%%%%%%%%%%%%%%%%%%%%

\section{Pulsation damping in the microphysical equation of state}
\label{appendix:pulsation_damping}

The damping of the post-migration pulsations observed in the models of
family UM with the microphysical EoS is much stronger than in family
US with the simple MUn EoS (cf.\
Figs.~\ref{fig:density_evolution_microphysical_eos}
and~\ref{fig:density_evolution_mun_eos}, respectively). It is caused
by a particular, very efficient damping mechanism that occurs if
matter moves across a first-order phase transition in such a type of
EoS due to quasi-radial pulsations \citep{bisnovatyi_84_a,
  haensel_89_a}. In the following, we explain that mechanism and the
relevant energy and time scales in our models in more detail. For
simplicity, without loss of generality we do not take into account
effects of rotation and only consider spherically symmetric Newtonian
models.

%%%%%%%%%%%%%%%%%%%%%%%%%%%%%%%%%%%%%%%%%%%%%%%%%%%%%%%%%%%%%%%%%%%%%%%%%%%%%%%%%%%
%%%%%%%%%%%%%%%%%%%%%%%%%%%%%%%%%%%%%%%%%%%%%%%%%%%%%%%%%%%%%%%%%%%%%%%%%%%%%%%%%%%

\subsection{Radial pulsations of the dense core}
\label{subappendix:radial_pulsations}

First we consider a rather simplified model of a pulsating compact
star with an EoS that exhibits a phase transition and leads to a dense
core in the center of the star. We neglect finite compressibility of
matter outside the core (except for the phase transition), so that the
star's density there is simply $ \rho_1 $.

Conservation of mass then leads to a relation between the pulsation
amplitude $ \delta r_\mathrm{c} $ of the dense core radius and the
radial velocity $ v_{r > r_\mathrm{c}} $ of the pulsations outside the
core,
\begin{equation}
  v_{r > r_\mathrm{c}} = \omega \, \delta r_\mathrm{c}
  \left( \frac{r_\mathrm{c}}{r} \right)^2 (\lambda - 1),
  \label{eq:dense_core_pulsation_amplitude}
\end{equation}
where $ r_\mathrm{c} $ is the unperturbed equilibrium radius of the
dense core, $ r $ is the unperturbed radial coordinate, $ \omega $ is
the angular pulsation frequency, and $ \lambda = \rho_2 / \rho_1 $
quantifies the jump in density at the core edge due to the phase
transition in the EoS. The phases of $ \delta r_\mathrm{c}(t) $ and
$ v_{r > r_\mathrm{c}} (t) $ differ by $ \pi $. As the core matter is
incompressible, the velocity $ v_{r < r_\mathrm{c}} $ at radii smaller
than $ r_\mathrm{c} $ vanishes.

Under our simple model assumptions, the average kinetic energy
contained in the radial pulsations is then given by
\begin{eqnarray}
  \overline{\!E}_\mathrm{p} & = &
  \pi \! \int_{r_\mathrm{c}}^{r_\mathrm{s}} \!\! dr \, r^2 \rho_1 \,
  v_{r > r_\mathrm{c}}^2 
  \nonumber
  \\
  & = & \pi \, \rho_1 \, r_\mathrm{c}^3 \, \omega^2 (\lambda - 1)^2
  \left( 1 - \frac{r_\mathrm{c}}{r_\mathrm{s}} \right)
  (\delta r_\mathrm{c})^2,
  \label{eq:pulsation_energy_from_dense_core_pulsation}
\end{eqnarray}%
where $ r_\mathrm{s} $ is the unperturbed equilibrium radius of the
star.

\begin{figure}
  \centerline{\includegraphics[width = 85 mm]{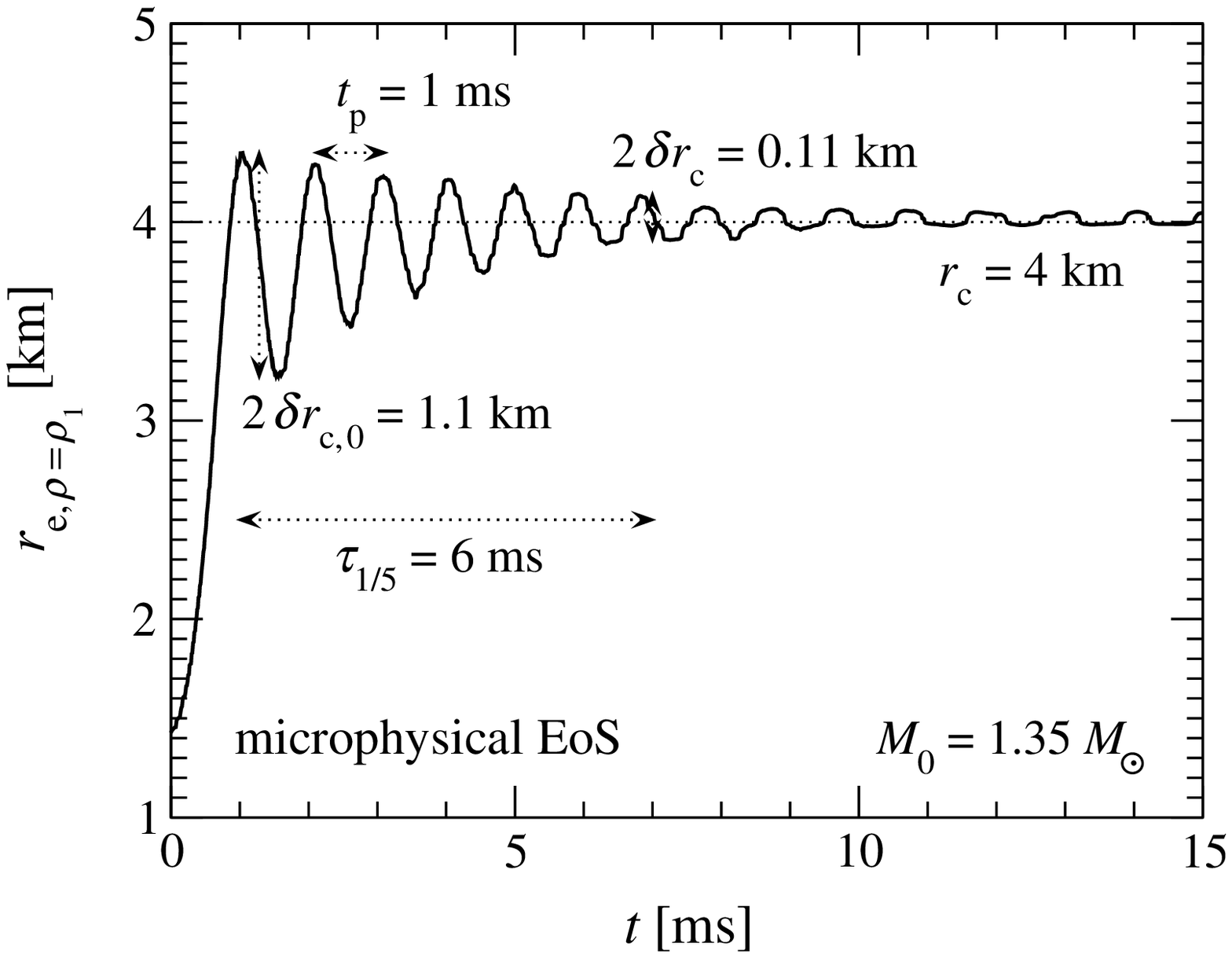}}
  \caption{Time evolution of the radius $ r_\mathrm{e,\rho = \rho_1} $
    of the dense core in the equatorial plane for the slowest rotating
    model UM1 of family UM with the microphysical EoS. The pulsation
    period of $ r_\mathrm{e,\rho = \rho_1} $ is roughly
    $ t_\mathrm{p} = 1 \mathrm{\ ms} $. The pulsation amplitude
    $ \delta r_\mathrm{c} $ is damped to $ 1 / 5 $ of its initial
    value of about $ \delta r_\mathrm{c,0} = 0.55 \mathrm{\ km} $ in
    approximately $ \tau_{1/5} = 6 \mathrm{\ ms} $. The equatorial
    radius of the dense core finally settles at the equilibrium value
    $ r_\mathrm{c} = 4 \mathrm{\ km} $.}
  \label{fig:dense_core_pulsations}
\end{figure}

The stable equilibrium initial model PMM1 that corresponds to the
post-migration state of model UM1 has a stellar radius
$ r_\mathrm{s,0} \approx 12 \mathrm{\ km} $, while from
Fig.~\ref{fig:dense_core_pulsations} we can extract typical
values $ \delta r_\mathrm{c,0} \approx 0.55 \mathrm{\ km} $ and
$ r_\mathrm{c,0} \approx 4 \mathrm{\ km} $ at the start of the
post-migration pulsations\footnote{For the other models of family UM
  we obtain similar values for $ \delta r_\mathrm{c,0} $ and
  $ r_\mathrm{c,0} $ and $ r_\mathrm{s,0} $. In
  Fig.~\ref{fig:dense_core_pulsations} we show the evolution of
  $ r_\mathrm{\rho = \rho_1} $ only in the equatorial plane, as both
  the qualitative behavior and the typical scale of this quantity
  does not depend much on latitude.}. For our microphysical EoS the
density discontinuity corresponds to $ \lambda = 1.62 $ (see
Section~\ref{subsection:equations_of_state} and
Fig.~\ref{fig:equation_of_state}). Thus, from
Eq.~(\ref{eq:pulsation_energy_from_dense_core_pulsation}) we obtain
for the initial pulsation energy:
\begin{equation}
  \overline{\!E}_\mathrm{p,0} \approx 3.7 \times 10^{48} \mathrm{\ erg}.
  \label{eq:initial_pulsation_energy_from_dense_core_pulsation}
\end{equation}

%%%%%%%%%%%%%%%%%%%%%%%%%%%%%%%%%%%%%%%%%%%%%%%%%%%%%%%%%%%%%%%%%%%%%%%%%%%%%%%%%%%
%%%%%%%%%%%%%%%%%%%%%%%%%%%%%%%%%%%%%%%%%%%%%%%%%%%%%%%%%%%%%%%%%%%%%%%%%%%%%%%%%%%

\subsection{Dissipation from a first-order phase transition and
  associated damping time scale}
\label{subappendix:phase_transition_dissipation}

We now consider the situation where the matter on both sides of
$ r_\mathrm{c} $ is fully equilibrated, i.e.\ realizes a minimum
of the appropriate thermodynamic potential \citep{bisnovatyi_84_a}.
The kinetic energy associated to the pulsations is dissipated during
the first-order phase transition, as matter passes through the
boundary of the dense core\footnote{Unlike in the estimate in
  Appendix~\ref{subappendix:radial_pulsations}, here the density
  outside the core need not be constant.}. Actually, the dissipation
occurs during the compression phase, which is the first half of the
pulsation period $ t_\mathrm{p} $, and takes place within the
shock front that forms at the dense core edge. In the thin layer
between the two phases, the sound speed $ c_\mathrm{s} \approx
[(P_2 - P_1) / (\mathcal{E}_2 - \mathcal{E}_1)]^{1/2} $
is rather low, because the large density jump is associated with only
a very small pressure increase. Therefore the matter flow within this
layer is supersonic. In reality, the dissipation within the shock
heats up the matter that flows into the dense core. However, in our
approximation we neglect thermal effects in the EoS, and thus
dissipation manifests itself only via the loss of kinetic energy in
the matter flow\footnote{As explained in
  Footnote~\ref{footnote:pressure_gradient}, in our simulations we
  introduce a small pressure gradient between the two phases, and thus
  the flow there remains subsonic. Nevertheless, the large density
  jump leads to numerical dissipation of comparable strength. We have
  also performed tests with almost zero pressure gradient and thus
  supersonic flow across the density jump, and find similar damping
  time scales of at most a factor 2 shorter. However, for numerical
  reasons here strong high-frequency noise superimposes the damped
  pulsations.}.

The average energy dissipated during one oscillation period is
\citep{bisnovatyi_84_a}
\begin{equation}
  \overline{\!E}_{\mathrm{d},t_\mathrm{p}} = \frac{4 \pi}{3} \,
  r_\mathrm{c}^2 \left( \frac{2 \pi}{t_\mathrm{p}} \right)^2 \!\! \rho_1 \,
  \lambda (\lambda - 1) (\delta r_\mathrm{c})^3.
  \label{eq:dissipation_energy_from_dense_core_pulsation}
\end{equation}
This dissipation results in a damping of pulsations and the decrease
of $ \delta r_\mathrm{c} $.

Assuming that all terms in
Eq.~(\ref{eq:dissipation_energy_from_dense_core_pulsation}) except
$ \delta r_\mathrm{c} $ are constant, we obtain for the average
dissipated power
\begin{equation}
  \dot{\overline{\!E}}_\mathrm{d} =
  \overline{\!E}_{\mathrm{d},t_\mathrm{p}} / t_\mathrm{p} =
  \dot{\overline{\!E}}_\mathrm{d,0}
  \left( \frac{\delta r_\mathrm{c}}{\delta r_\mathrm{c,0}} \right)^3\!\!\!.
  \label{eq:dissipated_power_from_dense_core_pulsation}
\end{equation}
Evaluating Eq.~(\ref{eq:dissipation_energy_from_dense_core_pulsation})
at the initial pulsation time and using the approximate value
$ t_\mathrm{p} \approx 1 \mathrm{\ ms} $ extracted from
Fig.~\ref{fig:dense_core_pulsations} together with the other estimates
from Appendix~\ref{subappendix:radial_pulsations} and $ \rho_1 $ for
the microphysical EoS, the power dissipated during the first pulsation
period is
\begin{equation}
  \dot{\overline{\!E}}_\mathrm{d,0} \approx
  3 \times 10^{48} \mathrm{\ erg\ ms}^{-1}\!.
  \label{eq:initial_dissipated_power_from_dense_core_pulsation}
\end{equation}

We now demand that the change in pulsation energy is entirely caused
by the above dissipation mechanism, and thus we arrive at
\begin{equation}
  \frac{d \overline{\!E}_\mathrm{p}}{dt} =
  - \dot{\overline{\!E}}_\mathrm{d}.
  \label{eq:pulsation_and_dissipation_power_relation}
\end{equation}
Rewriting Eq.~(\ref{eq:pulsation_energy_from_dense_core_pulsation})
for the average kinetic energy associated to the pulsations as
\begin{equation}
  \overline{\!E}_\mathrm{p} = \overline{\!E}_\mathrm{p,0}
  \left( \frac{\delta r_\mathrm{c}}{\delta r_\mathrm{c,0}} \right)^2\!\!\!
  \label{eq:pulsation_energy_from_dense_core_pulsation_modified}
\end{equation}
and using Eq.~(\ref{eq:dissipated_power_from_dense_core_pulsation}),
Eq.~(\ref{eq:pulsation_and_dissipation_power_relation}) can be
resolved for $ \delta r_\mathrm{c} $:
\begin{equation}
  \frac{1}{(\delta r_\mathrm{c})^3}
  \frac{d \, (\delta r_\mathrm{c})^2}{dt} =
  - \frac{1}{\delta r_\mathrm{c,0}} \,
  \frac{\dot{\overline{\!E}}_\mathrm{d,0}}{\overline{\!E}_\mathrm{p,0}}.
  \label{eq:expression_for_pulsation_amplitude}
\end{equation}
This can be straightforwardly integrated to
\begin{equation}
  \delta r_\mathrm{c} = \delta r_\mathrm{c,0}
  \left( 1 + \frac{\dot{\overline{\!E}}_\mathrm{d,0} \, t}
  {2 \, \overline{\!E}_\mathrm{p,0}} \right)^{-1} \!\!\! \approx
  \delta r_\mathrm{c,0} \, \frac{2 \, \overline{\!E}_\mathrm{p,0}}
  {\dot{\overline{\!E}}_\mathrm{d,0} \, t}.
  \label{eq:integration_of_pulsation_amplitude}
\end{equation}

With this we have an explicit expression for the time evolution of the
pulsation amplitude $ \delta r_\mathrm{c} $ of the dense core, which
we can apply to get an estimate for the damping time scale of the
pulsations. Choosing for instance $ t $ in
Eq.~(\ref{eq:integration_of_pulsation_amplitude}) as the time
$ \tau_{1/5} $ when the radial perturbations $ \delta r_\mathrm{c} $ of
the dense core have dropped to $ \delta r_\mathrm{c,0} / 5 $ and using
the values from
Eqs.~(\ref{eq:initial_pulsation_energy_from_dense_core_pulsation},
\ref{eq:initial_dissipated_power_from_dense_core_pulsation}) for
$ \overline{\!E}_\mathrm{p,0} $ and
$ \dot{\overline{\!E}}_\mathrm{d,0} $, respectively, we obtain
$ \tau_{1/5} \approx 12 \mathrm{\ ms} $. The value
$ \tau_{1/5} \approx 6 \mathrm{\ ms} $ read off from
Fig.~\ref{fig:dense_core_pulsations} is shorter, but of the same order
of magnitude ($ \sim 10 \mathrm{\ ms} $) as this crude analytic
estimate.

Although this treatment is somewhat simplistic, it demonstrates that
the compression of matter passing through the phase transition in the
microphysical EoS can very efficiently dissipate pulsation energy on a
time scale that is consistent with our observations for the family UM
of migration models. In contrast, as apparent from
Fig.~\ref{fig:density_evolution_mun_eos} the typical damping times of
the models of family US with the simple MUn EoS are much longer, as
there the above efficient dissipation mechanism is not at work.

%%%%%%%%%%%%%%%%%%%%%%%%%%%%%%%%%%%%%%%%%%%%%%%%%%%%%%%%%%%%%%%%%%%%%%%%%%%%%%%%%%%
%%%%%%%%%%%%%%%%%%%%%%%%%%%%%%%%%%%%%%%%%%%%%%%%%%%%%%%%%%%%%%%%%%%%%%%%%%%%%%%%%%%

\subsection{Non-equilibrium effects and bulk viscosity damping}
\label{subappendix:bulk_viscosity_damping}

We now relax the condition of complete equilibration. The phase
transition at the core radius $ r_\mathrm{c} $ must then be treated by
taking into account the slowness of weak interaction processes.
Consequently, kaon condensation will proceed off thermodynamic
equilibrium. However, the rate of non-equilibrium reactions increases
very steeply with the deviation from equilibrium. Therefore, also in
this case $ \dot{\overline{\!E}}_\mathrm{d} $ is not significantly
smaller than the upper bound obtained in
Appendix~\ref{subappendix:phase_transition_dissipation} under the
assumption of full equilibration.

In addition, non-equilibrium processes produce entropy in the form of
heating up the matter. This effect can be represented by bulk
viscosity, corresponding to an additional average dissipation power
$ \dot{\overline{\!E}}_\mathrm{d,bv} $.

In the case of a mixed quark-nucleon core, the efficient damping
mechanism described in
Section~\ref{subappendix:phase_transition_dissipation} does not work,
as there exists no single phase interface with a large jump in
density. Then the bulk viscosity generated by non-equilibrium,
strangeness changing reactions in the quark component,
\begin{equation}
  u + d \leftrightarrows s + u,
\end{equation}
becomes the main damping mechanism \citep[see][and references
therein]{madsen_99_a}.

If one considers the mini-collapse of an unstable neutron star caused
for instance by dynamic migration as in this work rather than a
supernova core collapse, the relative post-collapse pulsation
amplitude $ |\delta r_\mathrm{s}| / r_\mathrm{s} $ is typically less
than $ 6 \times 10^{-2} T / 10^{9} \mathrm{\ K} $. Therefore, bulk
viscosity does not depend on $ \delta r_\mathrm{s} $ and the
associated damping exhibits the usual form \citep{madsen_99_a}
\begin{equation}
  \delta r_\mathrm{s} =
  \delta r_\mathrm{s,0} \, e^{- t / \tau_\mathrm{bv,q}},
\end{equation}
where $ \tau_\mathrm{bv,q} $ is the damping time scale. For a strange
quark mass of $ 100 \mathrm{\ MeV} $, a Fermi energy of the down quark
of $ 300 \mathrm{\ MeV} $, and choosing $ \omega_\mathrm{p} =
2 \pi / t_\mathrm{p} \approx 2 \times 10^4 {\ s}^{-1} $ one gets,
after transforming a formula given in \citep{madsen_99_a},
\begin{equation}
  \tau_\mathrm{bv,q} \approx
  0.1 \, (T_9)^{-2} \left( \frac{r_\mathrm{c}}{0.3 \, r_\mathrm{s}} \right)^{-3}
  \!\!\! \mathrm{\ s}.
  \label{eq:bulk_viscosity_time_scale_quark_core}
\end{equation}

If the core, however, consists of a kaon condensate, then the bulk
viscosity is generated by the non-equilibrium processes
\begin{equation}
  n \leftrightarrows p + K^-
\end{equation}
that also change strangeness. For non-superfluid nucleons, this bulk
viscosity is independent of temperature and causes a damping of the
fundamental radial mode pulsations on a time scale
\citep{chatterjee_07_a}
\begin{equation}
  \tau_\mathrm{\,bv,k} \approx
  100 \left( \frac{r_\mathrm{c}}{0.3 \, r_\mathrm{s}} \right)^{-3}
  \!\!\! \mathrm{\ s},
  \label{eq:bulk_viscosity_time_scale_kaon_core}
\end{equation}
which is much longer than the corresponding bulk viscosity damping
time scale $ \tau_\mathrm{bv,q} $ in a quark core given in
Eq.~(\ref{eq:bulk_viscosity_time_scale_quark_core}). Thus here the
contribution to damping from bulk viscosity is only a very small part
of the total dissipation power, which is dominated by the effects from
the first-order phase transition.

\label{lastpage}

\end{document}